\documentclass[aps,11pt,twocolumn,prd,reprint,superscriptaddress,nofootinbib,floatfix,preprintnumbers]{revtex4-2}

\usepackage[utf8]{inputenc}
\usepackage{graphicx}
\usepackage{amsmath, amssymb}
\usepackage{mathtools}
\usepackage{physics}
\usepackage[english]{babel}
\usepackage[colorlinks=true, allcolors=blue]{hyperref}
\usepackage{url}
\usepackage{hyperref}
\usepackage{cleveref}
\Crefname{equation}{Eq.}{Eqs.}
\usepackage{xcolor}
\usepackage{tabularx}
\usepackage{soul}
\usepackage[lofdepth,lotdepth]{subfig}
\usepackage{footmisc}

\usepackage{lipsum}

\graphicspath{{./figures/}}

\def\calH{\mathcal{H}}
\def\Mpl{M_{\rm Pl}}
\renewcommand{\arraystretch}{1.5}
\begin{document}
\preprint{YITP-26-06}
\title{Constraining the inflaton potential with gravitational waves from oscillons}

\author{\textsc{Kaloian D. Lozanov}}
    \affiliation{Department of Applied Physics, Faculty of Applied Mathematics and Informatics,
Technical University of Sofia,
8, Saint Kliment Ohridski Blvd, Sofia 1000, Bulgaria}
    \affiliation{Asia Pacific Center for Theoretical Physics (APCTP), Pohang 37673, Korea}
    \affiliation{Kavli Institute for the Physics and Mathematics of the Universe (WPI), The University of Tokyo, Chiba 277-8583, Japan}
\author{\textsc{Misao Sasaki}}
    \affiliation{Asia Pacific Center for Theoretical Physics (APCTP), Pohang 37673, Korea}
    \affiliation{Kavli Institute for the Physics and Mathematics of the Universe (WPI), The University of Tokyo, Chiba 277-8583, Japan}
    \affiliation{Center for Gravitational Physics and Quantum Information, Yukawa Institute for Theoretical Physics, Kyoto University, Kyoto 606-8502, Japan}
    \affiliation{Leung Center for Cosmology and Particle Astrophysics, National Taiwan University, Taipei 10617, Taiwan}
\author{\textsc{Jan Tränkle}}
    \email{{jan.traenkle}@{itp.uni-hannover.de}}
    \affiliation{Institute for Theoretical Physics, Leibniz University Hannover, Appelstraße 2, 30167 Hannover, Germany}

\begin{abstract}
Under certain conditions, the oscillating inflaton condensate filling the Universe after inflation can fragment and form so-called oscillons. These long-lived soliton-like field configurations can dominate the Universe for several $e$-folds of expansion, leading to an early matter-dominated phase preceding the standard radiation era.
In this paper we show how the rapid final decay of the oscillons leads to an enhanced production of induced gravitational waves, whose energy density can saturate the observational bound on the effective number of relativistic species. We leverage this bound to constrain the inflaton mass, cubic, and quartic self-coupling in generic models that admit oscillon formation, providing novel and complementary constraints in regions of parameter space that are inaccessible with cosmic microwave background observations alone.
\end{abstract}

\maketitle

\section{Introduction}\label{sec:Introduction}
The idea of cosmic inflation was originally introduced to address questions not answered in the standard Big Bang model, like the absence of magnetic monopoles, or the flatness and horizon problems \cite{Starobinsky:1980te, Kazanas:1980tx, Guth:1980zm, Linde:1981mu, Albrecht:1982wi}.
Furthermore, its incarnation in terms of a quantum field slowly rolling down its potential offers a natural explanation for the origins of the spectrum of temperature anisotropies observed in the cosmic microwave background (CMB) \cite{Planck:2018vyg, Planck:2018jri, ACT:2025fju, ACT:2025tim} and the large scale structure of the Universe \cite{Mukhanov:1981xt, Kodama:1984ziu, Sasaki:1986hm, Mukhanov:1990me, Liddle:2000cg, Baumann:2009ds}.

Although the paradigm of single-field, slow-roll inflation has been successful in providing a unified solution to these problems, the nature of the hypothetical inflaton field and the precise shape of its potential are still unknown.
Furthermore, the dynamics during reheating, i.e.~the transition from inflation to the subsequent era of radiation domination, is largely unconstrained by current data. In fact, over a hundred different models for the inflaton potential have been proposed and there exists a large variety of different reheating expansion histories which are all compatible with current data \cite{Martin:2013tda, Allahverdi:2020bys}.
It is therefore crucial to look for observable signatures that might help distinguish between different models.

In this paper we focus on models in which the potential features a quadratic minimum which ``opens up" towards the wings. These properties are favored by observations, as many of the models that best fit the latest CMB data are precisely of this type \cite{Planck:2018jri, ACT:2025tim, Choudhury:2025vso}.
If the potential is shallower than quadratic, the inflaton condensate filling the Universe after inflation can fragment, and long-lived, non-linear field configurations known as oscillons can form via parametric self-resonance \cite{Copeland:1995fq, Kasuya:2002zs, Amin:2010jq, Amin:2010xe, Amin:2010dc, Amin:2011hj}.
They are localized, pseudo-stable, non-topological soliton-like objects, and can be thought of as spherical bubbles where the scalar field undergoes large amplitude oscillations on the inside. While inside the bubble the field probes the non-linear part of the potential, outside it oscillates around the quadratic minimum with small amplitude. Oscillons are not protected by any conserved charge, but they can persist for millions or more oscillations, which comes about as a dynamical balance between attractive self-interactions and dissipative forces. Their surprisingly long lifetimes and (classical and quantum) decay rates have been investigated both analytically and numerically, see e.g.~\cite{Kasuya:2002zs, Gleiser:2008ty, Hertzberg:2010yz, Salmi:2012ta, Saffin:2014yka, Kawasaki:2015vga, Ibe:2019vyo, Gleiser:2019rvw, Zhang:2020bec}.
Curiously, oscillons might also act as seeds for primordial black hole (PBH) formation \cite{Cotner:2018vug, Cotner:2019ykd}.

Importantly, the violent fragmentation of the inflaton condensate is accompanied by the generation of gravitational waves (GWs), which may serve as a direct probe of this elusive epoch of the Universe's history.
In particular, a stochastic GW background is sourced by the large inhomogeneities in the scalar field energy density generated during the preheating phase \cite{Khlebnikov:1997di, Easther:2006gt}, and the presence of oscillons may leave distinct imprints in the GW spectrum \cite{Zhou:2013tsa, Antusch:2016con, Antusch:2017vga, Lozanov:2019ylm, Hiramatsu:2020obh}.
Further, the formation of any cosmological solitons leads to a ``universal" gravitational wave signature due to the isocurvature perturbation associated with their number density fluctuations \cite{Lozanov:2023aez, Lozanov:2023knf}.

If oscillons do form abundantly and live long enough, they can come to dominate the post-inflationary universe and lead to an early matter-dominated era \cite{Amin:2011hj, Lozanov:2016hid, Lozanov:2017hjm, Jia:2024fmo}.
As their final decay to (scalar) radiation is almost instantaneous \cite{Olle:2019kbo, Zhang:2020bec}, such a scenario involves a sudden transition from matter domination to radiation domination, leading to an enhanced production of induced gravitational waves (GWs) through the so-called ``Poltergeist" mechanism \cite{Inomata:2019ivs}. Ref.~\cite{Lozanov:2022yoy} studied the enhanced production of GWs after oscillon decay from adiabatic curvature perturbations, assuming a scale-invariant power spectrum generated during inflation, while Ref.~\cite{Sui:2024grm} calculated the enhanced GWs due to the Poisson fluctuations of generic cosmological solitons.

In this paper, we compute the enhanced induced GWs due to the number density fluctuations of oscillons in several well-motivated inflationary models. By directly linking the GW signal to the parameters of the inflaton potential and applying the observational bound on the effective number of relativistic species $\Delta N_{\rm eff}$ to the induced GW amplitude, we demonstrate a novel way to constrain inflationary physics.
The structure of the paper is as follows. We give a general discussion of oscillon formation and their  properties in \Cref{sec:dynamics_after_inflation}, introducing also the specific models we consider in the following. In \Cref{sec:perturbations_oscillons_domination} we discuss the evolution of perturbations in the early matter-dominated phase driven by the domination of oscillons. The enhanced induced GWs sourced by these scalar perturbations are discussed in \Cref{sec:Induced_GWs}. Using these results, we derive novel constraints on the inflaton potential in \Cref{sec:iGW_constraints}. We close with a summary and conclusions in \Cref{sec:summary}.

\section{Non-linear dynamics after inflation}\label{sec:dynamics_after_inflation}
After slow-roll inflation ends, the inflaton field starts oscillating coherently about the minimum of its potential. If the inflaton couples (strongly) to other fields, this condensate eventually fragments and decays into daughter particles of those fields in a process known as preheating \cite{Bassett:2005xm, Lozanov:2019jxc, Garcia:2023dyf}. If instead the coupling to other fields is not too strong and the self-coupling of the field is large enough, the condensate may instead fragment into quasi-stable or transient lumps of the inflaton field itself \cite{Dymnikova:2000dy, Amin:2010xe, Amin:2010dc, Lozanov:2017hjm}.
As we are interested in studying the effects of oscillon formation, we will impose that the inflaton couples only weakly to standard model (SM) fields, in order to avoid rapid decay of the condensate to SM particles before any oscillons can form. Oscillon formation and decay in the presence of a coupling to additional matter fields has been studied in e.g.~\cite{Antusch:2015ziz, Shafi:2024jig, Li:2025ioq, Li:2025xtf}.
In the following section, we provide a short overview of properties of inflaton oscillons.

\subsection{Conditions for oscillon formation}\label{sec:conditions_oscillon_formation}
Generally speaking, oscillons can form from a scalar field of mass $m$ coherently oscillating in a potential of the form,
\begin{equation}
    V(\phi) = \frac{1}{2}m^2\phi^2 + V_{\rm nl}(\phi) \,,
\end{equation}
where the “non-linear" part of the potential $V_{\rm nl}$ has a minimum at the origin $V_{\rm nl}(0)=0$ and “opens up" towards the wings \cite{Kasuya:2002zs, Amin:2010jq, Amin:2010xe, Amin:2011hj}. This condition means $V_{\rm nl}(\phi)<0$ for some range of $\phi$, which entails an attractive self-interaction of the field, which can counterbalance dissipative effects and sustain the localized overdensity inside the oscillon.
The case of a minimum at a non-zero field value $\phi_0$ can be treated with a simple field-redefinition, $\phi\to \tilde\phi \coloneq \phi-\phi_0$.

For a broad class of models one can expand $V_{\rm nl}(\phi)$ near the minimum as
\begin{align}
    V_{\rm nl}(\phi \ll M) \approx  g \phi^3 - \lambda \phi^4 + \cdots,
    \label{eq:expansion_V(phi)}
\end{align}
with
\begin{align}
g=c_g \frac{m^2}{M\phantom{^2}} \quad \text{and} \quad \lambda=c_\lambda \frac{m^2}{M^2} \,,
    \label{eq:couplings_self_IA}
\end{align}
where $c_g$ and $c_\lambda$ are non-dimensional $\mathcal{O}(1)$ numbers which depend on the specific potential.
Here, $M$ denotes the scale where the potential turns over from the slow-roll regime at large $\phi$ to the quadratic regime near the minimum.
For potentials which are symmetric about the minimum, we have $c_g=0$. 
The condition $V_{\rm nl}(\phi)<0$ for having an attractive self interaction then translates to $\lambda>0$.
For asymmetric potentials, the non-linear part of the potential $V_{\rm nl}$ is negative only towards one side of the minimum. In the case of $c_g>0$, the attractive self-interaction arises from the asymmetric $\phi^3$-term, which becomes negative for $\phi-\phi_0<0$ rendering $V_{\rm nl}(\phi<\phi_0)<0$.

From \Cref{eq:couplings_self_IA} we see that, if $M$ becomes too large, we have $g,\lambda\to 0$ and $V(\phi)$ is effectively quadratic. In this case the attractive self-interaction, which is crucial for long-lived oscillons to form, is suppressed by inverse powers of $M$. This indicates an upper bound on $M$, and indeed from lattice simulations it is found that efficient oscillon formation requires $M^2\ll \Mpl^2$ \cite{Amin:2011hj,Lozanov:2017hjm}, see also \cite{Kim:2017duj,Kim:2021ipz} for analytical conditions for inflaton fragmentation.

However, $M$ should also not be too small, for the following reason.
Oscillons are produced from the backreaction of an unstable mode, $k_{un}\sim m$. If we wish to impose that the process be classical, this means that the occupation number of the unstable mode of the bosonic field $\phi$ should be large, $n(k_{un})\gg1$. Backreaction takes place when the energy density of the unstable mode becomes comparable to the background energy density, that is $\bar{\rho}\sim \rho(k_{un})\sim k_{un}^4 n(k_{un})$. Since $\bar{\rho}\sim m^2M^2$, imposing $n(k_{un})\gg1$ leads to the generic condition for classicality, $M^2\gg m^2$. This condition translates to an upper bound on the strength of the self-interaction, $\lambda \ll 1$.
In other words, in the regime of strong self-coupling the classical treatment breaks down and quantum backreaction has to be taken into account. The theory becomes non-perturbative in this regime, hence classical simulations of oscillons are no longer valid.

\subsection{Properties of individual oscillons}
The energy contained in a single lump of the scalar field, i.e.~the mass of a single oscillon, is determined by the inflaton mass and self-interaction as \cite{Amin:2019ums}
\begin{equation}
    m_{\rm osc} = \alpha_m \frac{M^2}{m\phantom{^2}} \,,
    \label{eq:m_osc}
\end{equation}
where $\alpha_m \sim \mathcal{O}(10^2 - 10^3)$ is a numerical coefficient that has to be determined from simulations. We will fix $\alpha_m=10^2$ as a benchmark for our plots.
This scaling relation is expected to be universal for large-amplitude oscillons, for which the amplitude at the core is ${\phi_{\rm core}\sim M}$ (see e.g.~\cite{Zhang:2020bec}). The relation \Cref{eq:m_osc} can be understood by noting that the radius for large-amplitude oscillons is approximately ${R_{\rm osc}\sim 0.1 m^{-1} }$ (or at least ${R_{\rm osc}\propto m^{-1}}$), implying that the mass of an individual oscillon scales as ${m_{\rm osc}\sim R_{\rm osc}^3\times V(\phi_{\rm core})\propto m^{-3} \times m^2\phi_{\rm core}^2}$ and thus we obtain ${m_{\rm osc}\propto M^2/m}$.

In the following we will assume the same mass $m_{\rm osc}$ given by \Cref{eq:m_osc} for all oscillons. This is justified because soon after their formation oscillons settle into a profile with a fixed frequency, where the decay rate is exponentially suppressed. This profile is an attractor and oscillons spend most of their lifetime in this state \cite{Zhang:2020bec}, rendering the oscillon mass function monochromatic.
%
\begin{figure*}[t]
\centering
\subfloat[Examples of potentials \labelcref{eq:Tanh-potential,eq:Monodromy-potential}.]{\includegraphics[width=0.49\textwidth]{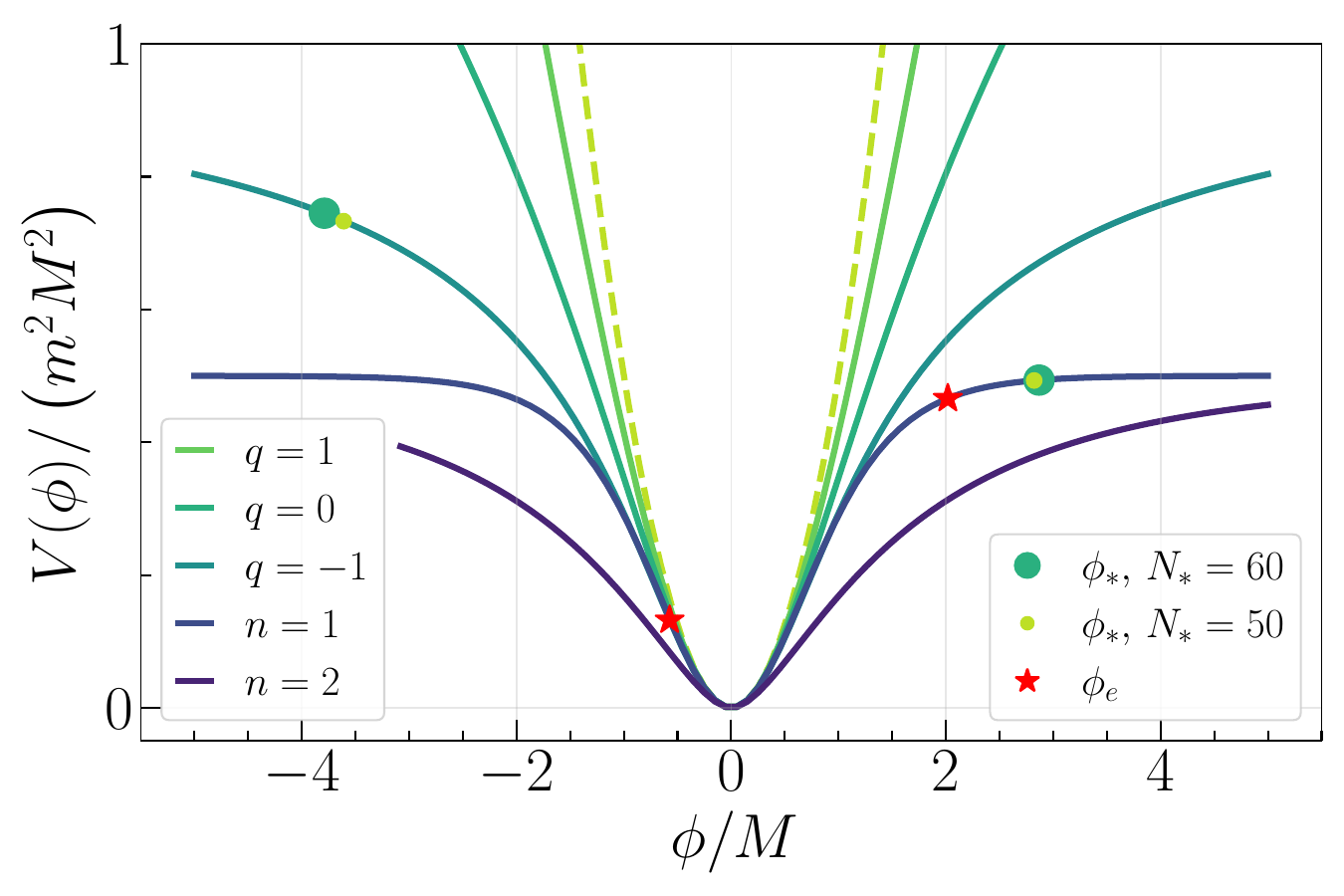}
\label{fig:potential_monodromy_attractors}}
\subfloat[Examples of potential \labelcref{eq:Hilltop-potential}.]{\includegraphics[width=0.49\textwidth]{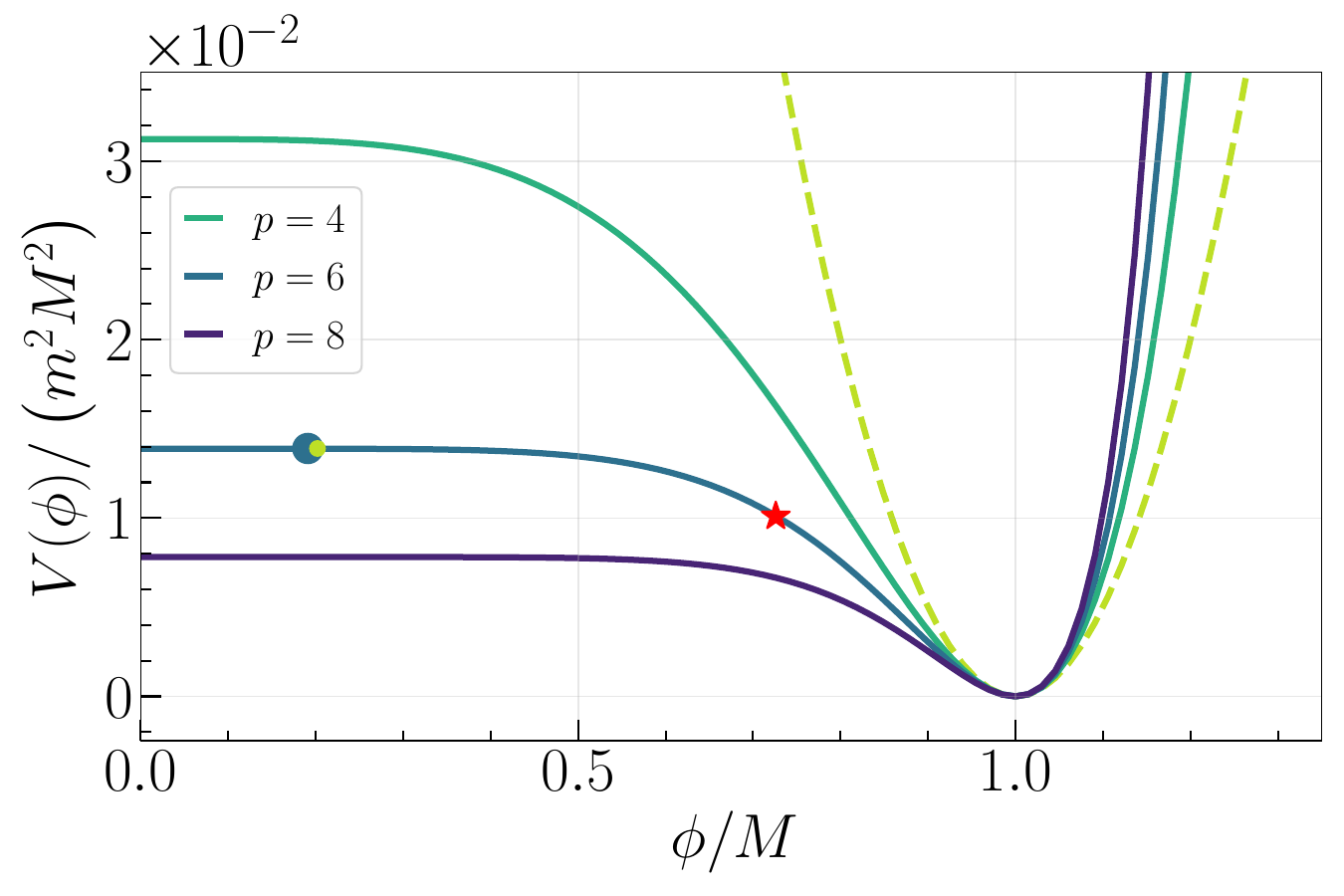}
\label{fig:potential_hilltop}}
\caption{Some representative examples of the potentials \labelcref{eq:Tanh-potential,eq:Monodromy-potential,eq:Hilltop-potential} for different values of $n$, $q$ and $p$, respectively. The dashed line marks the quadratic minimum in the two cases.
For a few examples, we also mark the field values $\phi_*$ and $\phi_e$, corresponding to scales that exit the horizon $N_*$ $e$-folds before the end of inflation (at $\phi_e$), with circles and stars, respectively.
In order to fix $\phi_*$ and $\phi_e$ for illustration purposes, we chose $M=0.1\Mpl$ for the T-model ($n=1$), and $M=2\Mpl$ for the plateau ($q=-1$) and $p=6$ cases.}
\label{fig:PotentialsPlot}
\end{figure*}
%

In numerical simulations it is found that oscillons typically persist for over a million oscillations
\begin{equation}
    \tau_{\rm osc} \gtrsim 10^6 m^{-1} \,,
    \label{eq:tau_osc}
\end{equation}
but the precise lifetime depends on the details of the potential and can in fact be much longer, exceeding $\tau_{\rm osc} \gtrsim 6 \times 10^8 m^{-1}$ in some cases. Due to their non-linear nature and the importance of backreaction effects, the exact lifetimes and decay rates have to be determined by numerical simulations, see e.g.~\cite{Gleiser:2008ty, Hertzberg:2010yz, Salmi:2012ta, Saffin:2014yka, Ibe:2019vyo, Gleiser:2019rvw, Zhang:2020bec} and in particular \cite{Lozanov:2017hjm, Olle:2019kbo, Antusch:2019qrr, Zhang:2020bec} for the models which we consider in the following.
Note that the precise lifetimes also depend on details like the initial oscillon radius and amplitude, and in some cases only lower bounds on the lifetime have been obtained. We will use the values provided in \cite{Lozanov:2017hjm, Olle:2019kbo, Antusch:2019qrr, Zhang:2020bec} and listed in \Cref{tab:comparison} below as benchmarks for our calculations.

\subsection{Inflationary potentials}\label{sec:Models}
%
\begin{table*}[t]
{\def\arraystretch{1.75}
\centering
    \begin{tabularx}{\linewidth}{|l|X|l|l|l|l|X|l|l|}
     \hline
     Model
     & $c_g$
     & $c_\lambda$
     & $r<0.056$
     & $r<10^{-3}$
     & $\tau_{\rm osc}=t_{\rm osc} m$
     & $N_{\rm osc}$
     & $N_*$
     & $\bar{H}_{\rm f}$ \\[3pt]
     \hline
     \hline
     
     $\alpha$-attractor T ($n=1$)
     & $0$
     & $1/3$
     & $M< 12\Mpl$
     & $M< 1.3\Mpl$
     & $\gtrsim 3\times 10^6$
     & $7$
     & $57$
     & $1/\sqrt{6}$ \\[3pt]
     \hline
     
     Axion monodromy ($q=1$) 
     & $0$
     & $1/8$
     & $M\lesssim\Mpl$\footnote{The bound on $r$ in the monodromy model implies $M\lesssim\Mpl$ and $N_*\gtrsim 71$. This corresponds to a rather large spectral index $n_s\approx 0.979$, which is disfavored by \textit{Planck}, but still within the error bars of the combined data with ACT \cite{ACT:2025fju}. \label{fn:footnote_AM}}
     & (ruled out)
     & $\sim 3\times 10^8$
     & $11$
     & $71$\footref{fn:footnote_AM}
     & $\sqrt{\frac{\Mpl}{3\sqrt{2} M}} $ \\[3pt]
     \hline
     
     Logarithmic ($q=0$)
     &$0$
     & $1/4$
     & $M< 3.2 \Mpl$
     & $M\ll \Mpl$
     & $\sim \mathcal{O}(1)\times 10^8$
     & $9$
     & $29-44$
     & $0.415$ \\[3pt]
     \hline
     
     Plateau ($q=-1$)
     & $0$
     & $3/8$
     & $M< 8.3 \Mpl$
     & $M< 0.018 \Mpl$
     & $\gtrsim 6\times 10^8$
     & $10$
     & $38-50$
     & $1/\sqrt{3}$ \\[3pt]
     \hline
     
     Hilltop ($p=4$)
     & $3/2$
     & $-17/8$
     & $M< 36 \Mpl$
     & $M< 8.5 \Mpl$
     & $\sim \mathcal{O}(1)\times10^5$
     & $4$
     & $54-86$
     & $1/(4\sqrt{2})$ \\[3pt]
     \hline
     
     Hilltop ($p=6$)
     & $5/2$
     & $-155/24$
     & $M< 51 \Mpl$
     & $M< 8.5 \Mpl$ 
     & $\sim \mathcal{O}(1)\times 10^6$
     & $5$
     & $53-71$
     & $1/(6\sqrt{2})$ \\[3pt]
     \hline
     
     Hilltop ($p=8$)
     & $7/2$
     & $-105/8$
     & $M< 67 \Mpl$
     & $M< 9.3 \Mpl$
     & $\sim 2\times 10^6$
     & $5$
     & $52-67$
     & $1/(8\sqrt{2})$ \\[3pt]
     \hline
    \end{tabularx}}
    \caption{Comparison of different models within the classes \Cref{eq:Tanh-potential,eq:Monodromy-potential,eq:Hilltop-potential} for different values of $n$, $q$ and $p$ for which results on oscillon lifetimes are available from numerical simulations, see \cite{Lozanov:2017hjm, Olle:2019kbo, Antusch:2019qrr, Zhang:2020bec}. We assume the central values of $\textit{Planck}$ \cite{Planck:2018vyg} for $A_s$, $n_s$ and the upper bound on $r$ to derive bounds on $M$ and the resulting range of $N_*$. We include bounds on $M$ that would be set by a future non-observation of $r$. The number of $e$-folds of oscillon domination $N_{\rm osc}$, cf.~\Cref{eq:N_osc}, is given for $M=10^{-2}\Mpl$. Note that recent ACT results \cite{ACT:2025fju} point to a somewhat higher value of $n_s$, which would slightly alter the numerical values. The Hubble parameter at oscillon formation is given in terms of $\bar{H}_{\rm f}=H_{\rm f} \Mpl/(m M)$ for ${M \ll \Mpl}$. }
    \label{tab:comparison}
\end{table*}
%
As stated in the introduction, there exists a plethora of models for inflation which satisfy current observational constraints \cite{Martin:2013tda}. We restrict ourselves to a minimal setup with a single inflaton field $\phi$ with canonical kinetic term. We consider different classes of potentials $V(\phi)$ with quadratic minima, which are theoretically well-motivated and among those that best fit current CMB data \cite{Martin:2013tda, Planck:2018jri, Choudhury:2025vso}.
Interestingly, the latest \textit{Planck} data \cite{Planck:2018vyg,Planck:2018jri} favors concave potentials, $V''(\phi_*)<0$, which indicates a constant plateau or power-law potential, $V(\phi\sim\phi_*)\sim\phi^{\alpha}$ with $\alpha\lesssim1$. This implies that attractive self-interactions, as described in \Cref{sec:conditions_oscillon_formation}, are present in the potential and inflation may naturally be followed by a phase of oscillon domination \cite{Amin:2011hj}. Here, $\phi_*$ denotes the value of the inflaton field around the time when CMB scales ${k_*\sim 10^{-3} \text{Mpc}^{-1}}$ exit the Hubble horizon.

The first family of models we consider is a hyperbolic tangent potential of the form,
\begin{equation}
    V(\phi) = \frac{m^2 M^2}{2} \tanh^{2n}\left( \left(\frac{|\phi|}{M}\right)^{1/n}\right) \,,
    \label{eq:Tanh-potential}
\end{equation}
which asymptotes to a plateau at large field values, $V(\phi\gg M)\approx m^2M^2/2$, for $n>0$.
This potential belongs to the $\alpha$-attractor T-models \cite{Kallosh:2013hoa, Kallosh:2013yoa}, motivated by (super-)conformal field theory and supergravity considerations. The parameter $M$ is related to the original $\alpha$-parameter via $M=\sqrt{6\alpha}\Mpl$.

Near the minimum, the non-linear part of the potential \Cref{eq:Tanh-potential} behaves like $V_{\rm nl}(\phi)\sim |\phi|^{2+2/n}$, indicating a more complex structure in the general case than captured by \Cref{eq:expansion_V(phi)}. We find that for $n=1$, the self-interaction is quartic with $c_g=0$ and $c_\lambda=1/3$. In the case $n=2$, the attractive self-interaction term is $|\phi|^3$ with $c_g=-2/3$, and $c_\lambda=-3/5$. Note that due to the absolute value of $\phi$ in \Cref{eq:Tanh-potential} the lowest order is cubic even though $V(\phi)$ is symmetric, and $c_g<0$.

The second class of potentials is given by
\begin{equation}
    V(\phi) = \frac{m^2 M^2}{q} \left(\left(1+\left(\frac{\phi}{M}\right)^2\right)^{q/2}-1\right) \,,
    \label{eq:Monodromy-potential}
\end{equation}
which is a generalization of the so-called axion monodromy models \cite{Silverstein:2008sg, McAllister:2008hb, McAllister:2014mpa}.
For positive values $q>0$, the potential \labelcref{eq:Monodromy-potential} has a power-law form at large field values, $V(\phi\gg M)\sim\phi^q$, while for negative $q<0$ it asymptotes to a constant $\sim m^2 M^2 / q$.
Near the minimum, the potential can be expanded as in \Cref{eq:expansion_V(phi)} with $c_g=0$ and ${c_\lambda = (2-q)/8}$. Intriguingly, the condition $q<2$, necessary to allow for the formation of oscillons, is imposed on the model by observational data \cite{Planck:2018jri}.

The last class we consider belongs to the so-called hilltop (squared) models of inflation, and the potential is of the form,
\begin{equation}
    V(\phi) = \frac{m^2M^2}{2 p^2 } \left(1-\left(\frac{\phi}{M}\right)^p\right)^2 \,,
    \label{eq:Hilltop-potential}
\end{equation}
which features a plateau (the “hilltop") near the origin, and a quadratic minimum around $\phi_0=M$.
Such models are theoretically attractive, as they appear commonly in particle physics models which involve a phase transition and (spontaneous) symmetry breaking \cite{Linde:1981mu, Albrecht:1982wi, Boubekeur:2005zm}.
In \Cref{eq:Hilltop-potential}, $p$ is a positive integer, and constraints from \textit{Planck} and recent ACT data favor $p\geq 4$ \cite{Planck:2018jri, Lynker:2025wyc}.
Expanding the potential in powers of $\tilde \phi =(\phi-M)\ll1$, we find $c_g=(p-1)/2$ and $c_\lambda = (-11+18p-7p^2)/24$.

The potentials in \Cref{eq:Tanh-potential,eq:Monodromy-potential} belong to the class of large-field models, where inflation happens at large field values $\phi > M$ in a plateau or power-law regime of the potential, whereas \Cref{eq:Hilltop-potential} represents a small field model, where inflation takes place with the field near the origin with $\phi < M$.
In \Cref{fig:PotentialsPlot} we illustrate some examples for potentials of the different classes \Cref{eq:Tanh-potential,eq:Monodromy-potential,eq:Hilltop-potential}, and in \Cref{tab:comparison} we list some representatives of each model for which numerical results for oscillon lifetimes are available.

Let us note that oscillons may also form in asymmetric potentials like the $\alpha$-attractor E-models \cite{Mahbub:2023faw, Lozanov:2017hjm}. This class of potentials includes the Starobinsky $R+R^2$ model \cite{Starobinsky:1980te} as a special case (for $\alpha=1$) \cite{Kallosh:2013yoa}.
Interestingly, as discussed above, for efficient oscillon production a sufficiently large self-interaction is necessary, which implies $\alpha\ll1$, such that in the Starobinsky model itself no significant oscillon formation is expected \cite{Takeda:2014qma, Mahbub:2023faw}.

The latest CMB data \cite{Planck:2018jri,Planck:2018vyg,ACT:2025fju,ACT:2025tim} provides constraints on the model parameters, in particular on the scale $M$. The procedure of relating parameters in the potential to CMB observables is illustrated in \Cref{app:parameter-relations}.
In all models we find that the current \textit{Planck} constraint, $r \lesssim 0.056$ \cite{Planck:2018vyg}, corresponds to an upper bound on $M$ which lies above the Planck scale, see \Cref{tab:comparison}. Future CMB experiments like LiteBIRD or the Simons Observatory are expected to reach sensitivities of $r \sim \mathcal{O}(10^{-3})$ \cite{SimonsObservatory:2018koc, LiteBIRD:2022cnt}, which could push the bound on $M$ to $\mathcal{O}(\Mpl)$. This means that a non-detection by these experiments, placing an upper bound $r < \mathcal{O}(10^{-3})$, would increasingly favor the parameter space where oscillon formation is generically expected.
This demonstrates that with growing experimental sensitivity, the non-trivial dynamics during reheating, involving e.g.~an oscillon-dominated epoch, have to be accounted for when assessing the physical viability of inflationary models.

\section{Perturbations in the oscillon-dominated universe}\label{sec:perturbations_oscillons_domination}
Let us now discuss the evolution of scalar perturbations after oscillons have formed. We provide some more details on the calculations in \Cref{app:isocurvature_scenario}.

In numerical simulations it is found that oscillons can come to dominate the Universe soon after their formation. Generically, once the scalar condensate fragments, over $\gtrsim50\%$ of the initial energy of the condensate is converted into oscillons, with the remaining energy density in scalar radiation \cite{Amin:2010xe, Amin:2010dc, Amin:2011hj}. As oscillons behave like non-relativistic matter, the subsequent expansion of the Universe makes them completely dominant. For example, in the $\alpha$-attractor T-model oscillons have been shown to become dominant within one $e$-fold after formation and lead to a matter-dominated phase, which can last several $e$-folds \cite{Lozanov:2016hid, Lozanov:2017hjm, Jia:2024fmo}.

For our analytical estimates, we assume that oscillon formation happens almost instantaneously at the time $t_{\rm f}$ when inflation ends,\footnote{In practice it takes a few oscillation periods for the condensate to fragment and form oscillons. However, because the oscillation frequency of the condensate, $\omega\sim\mathcal{O}(m)$, is much larger than the Hubble rate at the end of inflation, $H_{ e}\sim mM/\Mpl$, given that $M\ll \Mpl$, oscillon formation typically completes within less than one $e$-fold after inflation has ended \cite{Amin:2010dc,Lozanov:2017hjm,Amin:2019ums}, validating our assumption $H_{\rm e}\simeq H_{\rm f}$.}
as determined by the condition on the slow-roll parameter $\epsilon_V\simeq 1$ given in Eq.~\eqref{eq:slow-roll-eps}.
After they have formed, the population of oscillons can be treated as a dust-like perfect fluid with characteristic density fluctuations due to their discrete nature.
Assuming the formation process in real space can be considered as a set of uncorrelated localized events, the initial power spectrum of number density fluctuations is given by a Poisson distribution.
The assumption of an initially Poissonian spectrum is directly supported by lattice simulations \cite{Amin:2019ums}. These simulations show that the specific oscillon formation sites are spatially uncorrelated, resulting in a characteristic white-noise spectrum. The physical mechanism behind this is that fragmentation is driven by the local, self-resonant amplification of random vacuum fluctuations. At late times, gravitational interactions between these localized individual objects lead to clustering and modify the power spectrum on small scales. We account for this gravitational growth of density fluctuations in our analysis of cosmological perturbations as described in \Cref{app:isocurvature_scenario}. On scales much larger than the mean separation of individual oscillons, the local average density is additionally modulated by the nearly scale-invariant primordial fluctuations, reflecting the long-range correlations of the condensate. However, because the induced GW signal we will discuss in \Cref{sec:Induced_GWs} is entirely dominated by the small-scale Poisson fluctuations of the oscillon gas, we can safely neglect the large-scale tail of the power spectrum.

For the density contrast $\delta_{\rm osc}= \delta\rho_{\rm osc} / \rho_{\rm osc}$ (where $\rho_{\rm osc}$ and $\delta\rho_{\rm osc}$ are the energy density in oscillons and its perturbation, respectively), the power spectrum\footnote{The dimensionless power spectrum of a field $X$ is defined by
\begin{equation}
    \mathcal{P}_{X}(k) \delta_D(k+k')=\frac{k^3}{2\pi^2}\langle X_{\textbf{k}} X_{\textbf{k}'} \rangle \,.
\end{equation}} is given by \cite{Papanikolaou:2020qtd}
\begin{equation}
    \label{eq:Posc_Poiss}
    \mathcal{P}_{\delta_{\rm osc, f}}(k) = \frac{2}{3 \pi} \left(\frac{k}{k_{\rm osc}}\right)^3 \,,
\end{equation}
which is valid on scales $k\leq k_{\rm osc}$ and at the formation time denoted by the subscript ``f". Here, $k$ denotes the wavenumber of a given Fourier mode, and the scale $k_{\rm osc}$ corresponds to the comoving mean separation of individual oscillons. On smaller scales, $k > k_{\rm osc}$, the effective fluid description breaks down. 
Note that many oscillons form in each Hubble patch, such that $k_{\rm osc} > k_{\rm f}$ with the scale $k_{\rm f}=a_{\rm f}H_{\rm f}$ corresponding to the horizon size at formation.\footnote{Note that this is opposite to the case of PBHs, where only in a few Hubble patches a single PBH is formed and thus ${k_{\rm PBH} < k_{\rm f}}$ for the comoving mean PBH separation $k_{\rm PBH}$.} Here, $a$ denotes the scale factor of the Universe, and $H$ is the Hubble parameter.

Once oscillons become dominant, the total energy density fluctuation is dominated by the density fluctuations in the  oscillon fluid. Then, the Poisson equation for the gravitational potential $\Phi$ on subhorizon scales ($k \gg a H$) reads
\begin{align}
    \Phi \simeq \frac{3}{2} \left(\frac{aH}{k}\right)^2 \delta_{\rm osc} \,.
    \label{eq:Poisson_eq}
\end{align}
During matter domination $\Phi$ becomes constant, while the density contrast grows linearly with the scale factor, $\delta_{\rm osc}\propto a$ \cite{Baumann:2022mni}.
Let us emphasize here that the oscillon number density fluctuations correspond to isocurvature perturbations, meaning that initially the curvature perturbation is zero. However, numerically one can check that a curvature perturbation is quickly generated on super- and subhorizon scales for isocurvature initial conditions when $\beta\coloneq\rho_{\rm osc,f}/\rho_{\rm f}\gtrsim1/2$, see \Cref{fig:Phi_iso_plot} in \Cref{app:isocurvature_scenario}.

Deep in the matter dominated phase, the constant plateau value $\Phi_{\rm MD}\equiv\Phi(a\gg a_{\rm f})$ is determined by the initial amplitude of the isocurvature perturbation $S_{\rm f}$, as well as the initial energy fraction $\beta$.
As explained in \Cref{app:isocurvature_scenario}, we find a numerical fit for $\Phi_{\rm MD}$ by evolving the coupled equations \Cref{eq:PhiEq,eq:SEq} for $\Phi$ and $S$ deep into the matter-dominated phase. We obtain
\begin{align}
    \Phi_{\rm MD} & \approx S_{\rm f} \times \left( \frac{5 (4-\beta)}{4 (1-\beta)}+ \frac{1}{\mathcal{S}(\beta)} \left(\frac{k}{k_{\rm f}}\right)^2\right)^{-1} \,. \label{eq:PhiIsoeMDInterpol}
\end{align}
In the $k\gg k_{\rm f}$ regime $\Phi_{\rm MD}$ scales as $k^{-2}$, cf.~\Cref{eq:Phi_MD_Sub}, which follows directly from the Poisson equation \labelcref{eq:Poisson_eq}. In the superhorizon regime ($k\ll k_{\rm f}$) the scale-dependence vanishes, see \Cref{eq:Phi_MD_Super}. The suppression factor $\mathcal{S}(\beta)$ quantifies the efficiency of the conversion of isocurvature to curvature on subhorizon scales in dependence of the initial energy fraction of the matter component $\beta$. For $1/2\lesssim\beta<1$ it is of order $\mathcal{S}(\beta)\sim \mathcal{O}(10^{-1}-10^{-2})$, and we provide a numerical fit in \Cref{eq:SuppressionFactor}.
The initial amplitude of the isocurvature is proportional to the oscillon number density fluctuations, $S_{\rm f}\propto\delta_{\rm osc}$, and given explicitly in \Cref{eq:S_f}.

Let us focus on the regime where $k\sim k_{\rm osc} > k_{\rm f}$, which gives the largest contribution to the induced GWs. Combining \Cref{eq:PhiIsoeMDInterpol,eq:Posc_Poiss,eq:S_f} we then find the power spectrum of the curvature perturbation in the matter-dominated epoch as
\begin{align}
    \mathcal{P}_{\Phi_{\rm MD}}(k\gg k_{\rm f}) \approx \mathcal{C}(\beta)\frac{2}{3\pi} \left(\frac{k_{\rm f}}{k_{\rm osc}}\right)^4 \left(\frac{k}{k_{\rm osc}}\right)^{-1} \,.
    \label{eq:PowerSpectrumPhi}
\end{align}
The $\beta$-dependent prefactor is defined in \Cref{eq:def_beta} and is of order $\mathcal{C}(\beta)\sim \mathcal{O}(10^{-1})$ for our values of interest.

Let us mention that the (nearly) scale-invariant curvature perturbation generated during inflation could be accounted for by adding a constant term $\sim A_s$ to \Cref{eq:PowerSpectrumPhi}. This term could become relevant on very large scales, $k\ll k_{\rm f}$. The enhanced generation of induced GWs from such a flat spectrum in a similar setup has been studied in \cite{Lozanov:2022yoy}.

We can relate the relevant scales in the system to each other to find
\begin{equation}
    \frac{k_{\rm osc}}{k_{\rm f}} = \left(4 \pi \beta \frac{\Mpl}{m_{\rm osc}} \frac{\Mpl}{H_{\rm f}}\right)^{1/3} \,,
    \label{eq:kosc_kf}
\end{equation}
and
\begin{align}
    \frac{k_{\rm f}}{k_{\rm rh}} \approx \sqrt{\beta}\, \left( 1 + \frac{3}{2} t_{\rm osc} H_{\rm f}\right)^{1/3} \,,
    \label{eq:kf_krh}
\end{align}
where in the second equality we assumed $\beta>1/2$ and approximated the background evolution to be that of a purely matter-dominated universe throughout, i.e.~$a_{\rm MD}\propto t^{2/3}$. $t_{\rm osc}=\tau_{\rm osc}m^{-1}$ denotes the lifetime of the oscillons, and $k_{\rm rh}$ is the scale corresponding to the horizon size at the time of reheating, i.e.~when the oscillons decay into radiation.

\Cref{eq:kf_krh} also provides an expression for the number of e-folds from oscillon formation to decay, namely,
\begin{align}
    N_{\rm osc} &= 2 \ln\left(\frac{k_{\rm f}}{k_{\rm rh}}\right)-\ln(\beta)\nonumber \\
    &\approx \frac{2}{3} \ln\left( 1 + \frac{3}{2} t_{\rm osc} H_{\rm f}\right) \,.
    \label{eq:N_osc}
\end{align}
Inserting some typical numbers, as those provided in \Cref{tab:comparison}, we find that oscillons can persist for up to $\mathcal{O}(10)$ $e$-folds, depending on the values of the lifetime $t_{\rm osc}$ and $M/\Mpl$.
Let us note here that during such a matter-dominated phase density perturbations grow and can become non-linear, invalidating the use of linear cosmological perturbation theory. We provide an estimate of the scale $k_{\rm nl}$, beyond which non-linearities become relevant during the oscillon lifetime, in \Cref{app:k_NL}.

Using \Cref{eq:N_osc} we can also illustrate the timescale on which the oscillon decay takes place. For example, in the hilltop model the decay happens on a timescale $\Delta \tau_{\rm dec} m\sim 10^3$, compared to a total lifetime of $\tau_{\rm osc}m \sim 10^6$ for $M\sim 10^{-2}\Mpl$ \cite{Antusch:2019qrr}.
Using these values, the number of $e$-folds from oscillon formation to decay is around $N_{\rm osc}\sim 5$, while the final decay happens within $\Delta N_{\rm dec}\sim 10^{-3}$ $e$-folds, which shows that the approximation of instantaneous decay is well justified.

Finally, the temperature of the radiation bath after oscillon decay (i.e.~the reheating temperature)\footnote{Note that oscillons decay into scalar radiation, so we require a (weak) coupling of the inflaton to SM fields to allow the scalar radiation to decay into the SM radiation bath later.}
is given by
\begin{equation}
    T_{\rm rh} \simeq \left( \sqrt{\frac{90}{g_*(T_{\rm rh})\pi^2}} \frac{\beta^{-5/2} H_{\rm f} \Mpl}{1+\frac{3}{2}H_{\rm f}t_{\rm osc}}\right)^{1/2} \,,
    \label{eq:ReheatingTemp}
\end{equation}
where $g_*(T)$ denotes the number of relativistic degrees of freedom at a given temperature $T$.

\section{Gravitational waves induced by oscillon decay}\label{sec:Induced_GWs}
Large scalar density perturbations, such as the ones associated with the oscillon number density perturbations described above, can induce tensor perturbations, i.e.~gravitational waves, at second order in cosmological perturbation theory \cite{Tomita:1967wkp, Matarrese:1992rp, Ananda:2006af, Baumann:2007zm, Kohri:2018awv}, see \cite{Domenech:2021ztg} for a recent review.
The sudden transition from oscillon (i.e., matter) domination to radiation domination leads to an enhanced production of induced gravitational waves via the so-called ``poltergeist" mechanism \cite{Inomata:2019ivs, Inomata:2025wiv}. Physically, this can be understood as a consequence of the fact that during matter domination the sound speed of perturbations vanishes, $c_s\simeq 0$, while at the same time density perturbations grow linearly with the scale factor. After the sudden transition, the overdensities are converted into large perturbations in the radiation fluid, which start propagating with $c_s^2=1/3$, leading to large anisotropies and velocity flows, which backreact onto the metric to produce a loud GW signal.\footnote{Let us note here that in some models, oscillons decay into scalar particles with momenta comparable to the rest mass $m$ \cite{Imagawa:2021sxt}. Such quasi-relativistic particles would give rise to a fluid with a softer equation of state, $0<w<1/3$.
However, extending the formalism to this case would require a dedicated re-analysis of the induced GW kernel $\overline{\mathcal{I}^2}$, which enters the tensor power spectrum through \Cref{eq:TensorPowerSpectrum}, for the Poltergeist mechanism with the curvature transfer function for a generic perfect fluid background, which is beyond the scope of the present paper. As a rough order-of-magnitude estimate, we note that the $c_s$-dependent prefactor of \Cref{eq:Omega_GW_peak} changes by $\lesssim12\%$ when varying $w$ between $0.1\lesssim w\lesssim 1/3$, indicating that the mechanism is robust against small deviations of $w$ from the exact radiation case $w=1/3$.}

A very similar situation appears in the PBH reheating scenario, where ultra-light PBHs temporarily dominate the early Universe before evaporating and reheating the Universe via their Hawking radiation.\footnote{A crucial difference between the oscillon and PBH scenarios is that the initial energy density fraction of oscillons at formation is $\Omega_{\rm osc, f}\sim \mathcal{O}(1)$, compared to the PBH case where the initial fraction is constrained to be $\Omega_{\rm PBH, f}\ll 1$ from induced GW overproduction \cite{Papanikolaou:2020qtd, Domenech:2020ssp, Domenech:2024wao}. A further difference lies in the details of the decay mechanism, which in the case of PBHs proceeds through the (almost) thermal emission of Hawking radiation. Oscillons on the other hand spend most of their lifetime in a quasi-stable attractor configuration, oscillating at a frequency $\omega^2<m^2$. Once $\omega$ increases above a critical value $\omega_{\rm crit}$, this configuration ceases to be a solution to the non-linear equations of motion and the oscillon becomes unstable and rapidly dissipates \cite{Zhang:2020bec}.}
The enhanced induced GWs in this scenario have been first studied in \cite{Inomata:2020lmk, Papanikolaou:2020qtd, Domenech:2020ssp, Domenech:2021wkk} and in the following we adopt the notation and formalism developed in \cite{Domenech:2020ssp, Domenech:2024wao}. We provide some more details on the calculation in \Cref{app:SIGW}.

The dominant resonant contribution to the GW spectrum is obtained as
\begin{align} \label{eq:Omega_GW_res}
    \Omega_{\rm GW, res} (k) \approx \Omega_{\rm GW, res}^{\rm peak} \left(\frac{k}{k_{\rm osc}}\right)^5 \Theta_{\rm uv}(k) \,,
\end{align}
where the peak amplitude is given by
\begin{align}
    \Omega_{\rm GW, res}^{\rm peak} = \frac{\mathcal{C}(\beta)^2}{3456 \pi}  c_s(1-c_s^2)^2 \left(\frac{k_{\rm f}}{k_{\rm rh}}\right)^7 \left(\frac{k_{\rm f}}{k_{\rm osc}}\right) \,.
\end{align}
The function $\Theta_{\rm uv}(k)$, defined in \Cref{eq:Theta_UV}, acts as a smooth cut-off for frequencies above the UV scale $k_{\rm osc}$.
Using the relations \Cref{eq:kf_krh,eq:kosc_kf} we can rewrite the peak amplitude of the GW spectrum as
\begin{align}
    \Omega_{\rm GW, res}^{\rm peak} = & \frac{\mathcal{C}(\beta)^2 \beta^{19/6}}{3456 \pi}  c_s(1-c_s^2)^2 \left( 1 + \frac{3}{2} t_{\rm osc} H_{\rm f}\right)^{7/3} \nonumber \\ 
    &\times \left(\frac{1}{4\pi}\frac{m_{\rm osc}}{\Mpl}\frac{H_{\rm f}}{\Mpl}\right)^{1/3} \,,
    \label{eq:Omega_GW_peak}
\end{align}
in terms of the oscillon mass, lifetime and time of formation. In principle, these values are fixed for a given inflationary model and thus one can use this GW spectrum to directly probe the parameters of the inflaton potential.
Inserting the expression \Cref{eq:m_osc} for $m_{\rm osc}$ and those for $H_{\rm f}$ and $t_{\rm osc}$ given in \Cref{tab:comparison}, the dependence on $m$ cancels out and the amplitude of the spectrum is solely determined by the scale $M$ and the initial energy fraction $\beta$.

The low-frequency tail at $k\ll k_{\rm osc}$ scales linearly with $k$ and is given by
\begin{align}
    \Omega_{\rm GW, IR}(k) \approx \frac{c_s^4 \, \mathcal{C}(\beta)^2}{324 \pi^2}  \left(\frac{k_{\rm f}}{k_{\rm osc}}\right)^8 \left(\frac{k_{\rm osc}}{k_{\rm rh}}\right)^6 \frac{k}{k_{\rm osc}} \,.
    \label{eq:Omega_GW_IR}
\end{align}
Using that the GW energy density redshifts as that of radiation one can directly relate the GW spectral density in the radiation epoch to the spectral energy density today, as explicitly given in \Cref{eq:Omega_GW_0}.

The peak frequency ${f_{\rm peak}=k_{\rm osc}/(2\pi)}$ of the spectrum today is given by
\begin{align}
    f_{\rm peak}
    = &\frac{g_*(T_{\rm rh})^{1/4}}{2^{7/12}5^{1/4}\sqrt{3}\pi^{1/6}}\left(\frac{g_{*,s}(T_0)}{g_{*,s}(T_{\rm rh})}\right)^{1/3} T_0 \nonumber \\
    &\times \left(\frac{H_{\rm f}\Mpl \beta^{-5/2}}{m_{\rm osc}^2\left(1+\frac{3}{2}t_{\rm osc}H_{\rm f}\right)}\right)^{1/6} \,, \label{eq:f_peak}
\end{align}
where we used the relations \Cref{eq:kosc_kf,eq:kf_krh,eq:ReheatingTemp}. To redshift the frequency to today we used entropy conservation, ${a T g_{*,s}(T)^{1/3}=\text{const.}}$, with the number of entropic degrees of freedom $g_{*,s}(T)$ \cite{Baumann:2022mni}.
%
\begin{figure}
\centering
\includegraphics[width=0.48\textwidth]{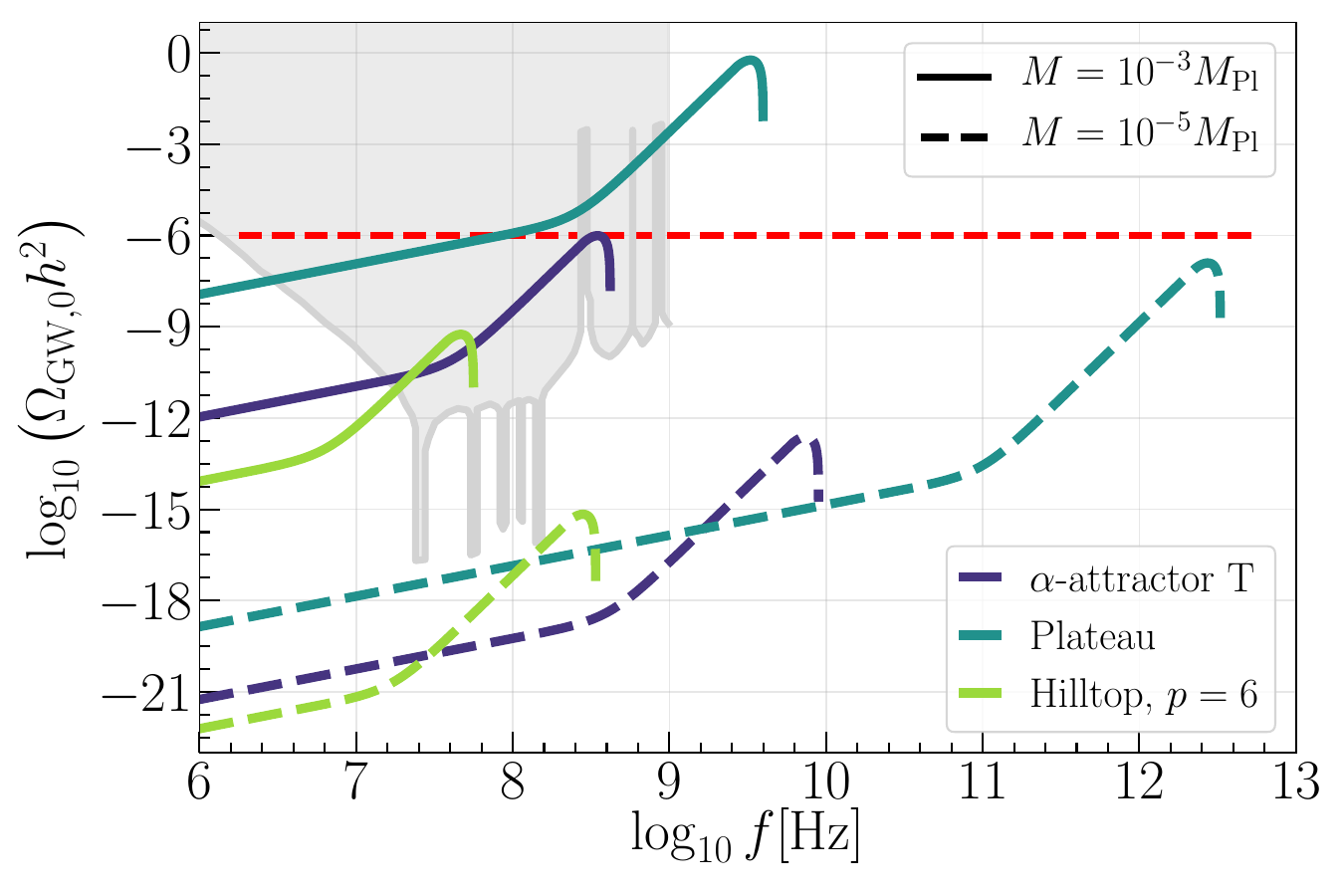}
\caption{Present day spectral energy density of gravitational waves $\Omega_{\rm GW,0}(f)$ induced after oscillon decay for some representative models. We fixed $\beta=0.8$ and two illustrative values of $M$ shown with solid and dashed curves, respectively. The horizontal red dashed line marks the (integrated) BBN bound on $\Delta N_{\rm eff}$, and the light gray curve indicates the projected sensitivity of resonant cavities \cite{Herman:2022fau}. As a reference, the reheating temperatures corresponding to the displayed models all lie in the range $9\times 10^{10}\lesssim T_{\rm rh}\left[ \text{GeV}\right] \lesssim 1\times 10^{13}$.
}
\label{fig:Omega_GW_0}
\end{figure}
%

We plot the present day GW spectrum for some sample values of $M$ and different potentials in \Cref{fig:Omega_GW_0}. See how for larger values of $M$ the peak amplitude increases, in some cases breaking the bound from Big Bang Nucleosynthesis (BBN) (as discussed below).
Inserting some benchmark values into \Cref{eq:f_peak}, we find that the peak frequency today is of the order $f_{\rm peak}\gtrsim \mathcal{O}(10^7)\rm Hz$, and could thus be a target for future high frequency GW detectors such as resonant cavities, re-purposed axion experiments using the inverse Gertsenshtein effect, or next-generation laser interferometers \cite{Aggarwal:2020olq, Berlin:2021txa, Herman:2022fau, Domcke:2022rgu, He:2023xoh}.
Interestingly, the enhanced induced GWs associated with the adiabatic, (nearly) scale-invariant spectrum of perturbations from inflation peak at slightly lower frequencies around $\sim k_{\rm f}$ and could therefore serve as a complementary probe \cite{Lozanov:2022yoy, Domenech:2024wao}.
Note also how the peak frequency shifts towards higher values upon decreasing $M$. Unfortunately, this indicates that in large regions of the parameter space compatible with BBN the peak frequency falls above $\mathcal{O}(\rm GHz)$, making a direct detection extremely challenging.
Let us mention here also that the sensitivity curve estimated in \cite{Herman:2022fau} might be regarded as very optimistic, and the sensitivity of resonant cavities to stochastic GW backgrounds (as compared to coherent, monochromatic signals) might be significantly reduced \cite{Berlin:2021txa, Aggarwal:2025noe}.

\section{Induced GW constraints on the inflaton potential}\label{sec:iGW_constraints}
%
\begin{figure*}[ht]
\centering
\subfloat[Parameter space of potential \labelcref{eq:Hilltop-potential}]{
\includegraphics[width=0.48\textwidth]{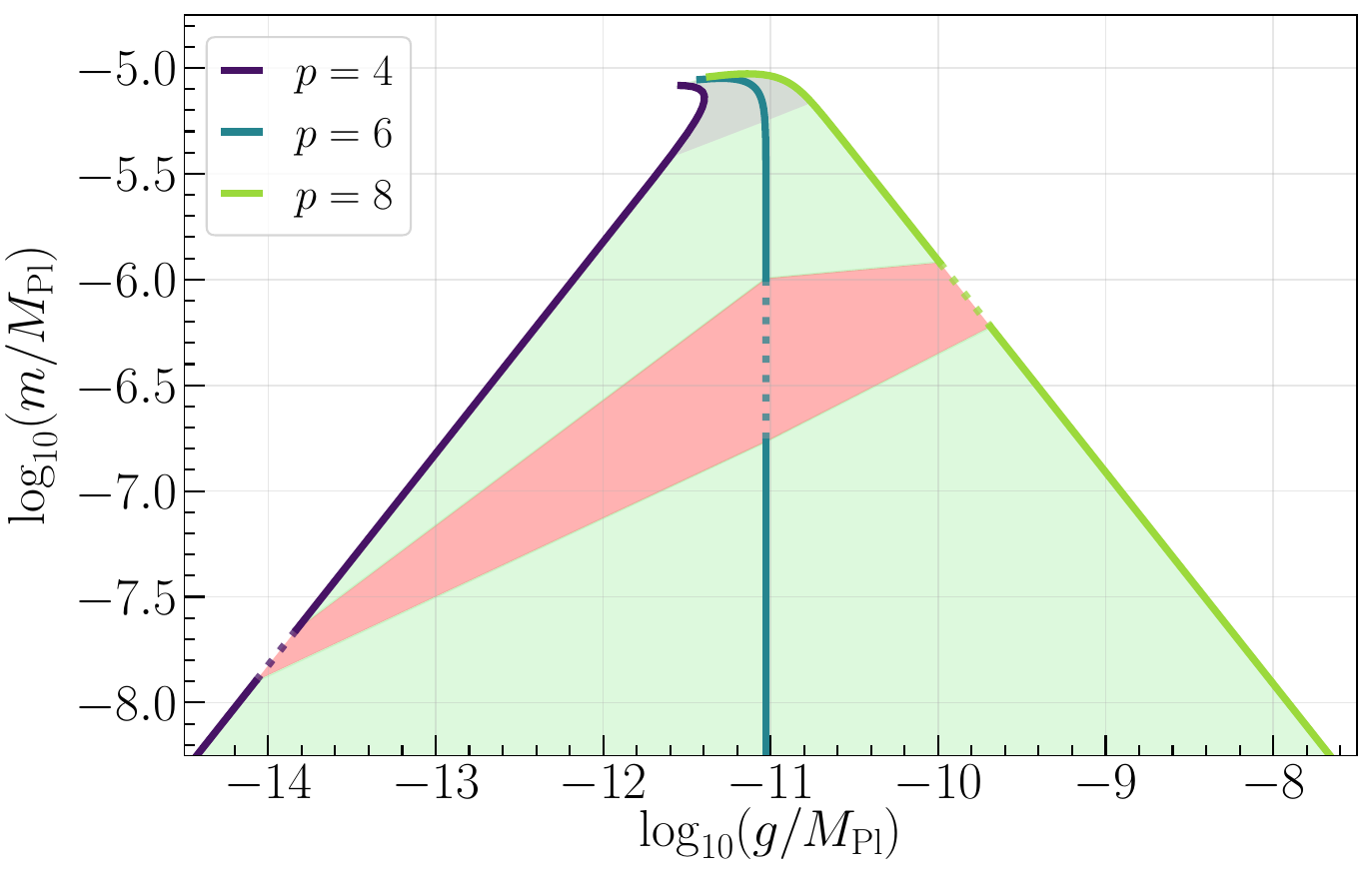}
\label{subfig:parameter_space_g}}
\subfloat[Parameter space of potentials \labelcref{eq:Tanh-potential,eq:Monodromy-potential}]{
\includegraphics[width=0.48\textwidth]{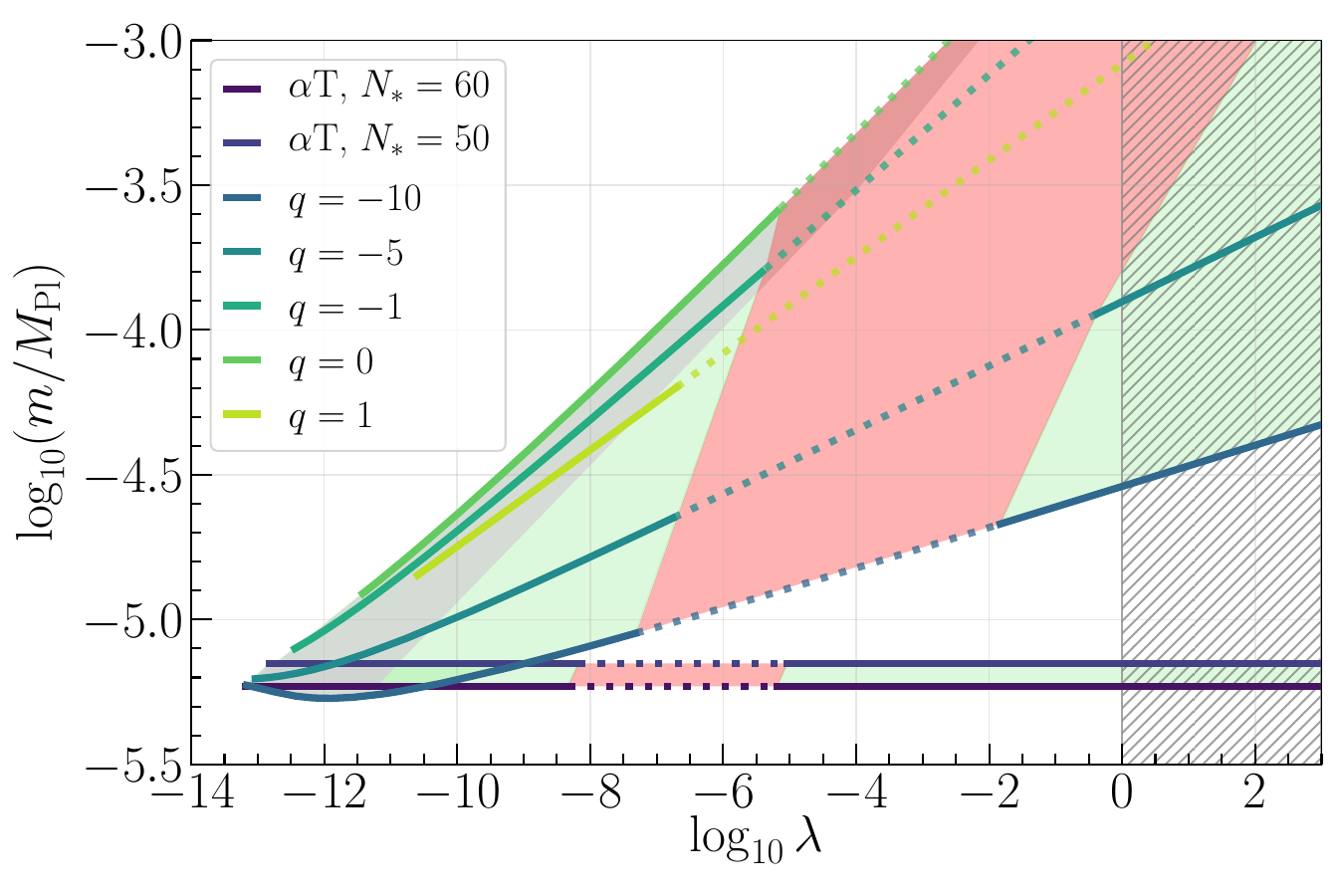}
\label{subfig:parameter_space_lam}}
\caption{Here we show the allowed and excluded parameter space in the $(m,\lambda)$-plane for the symmetric potentials \Cref{eq:Tanh-potential,eq:Monodromy-potential} ($\alpha$-attractor T-model and generalized monodromy), and in the $(m,g)$-plane for the asymmetric hilltop potential \Cref{eq:Hilltop-potential}. Each line marks one specific model. The green shaded area is allowed parameter space, the red shaded region is ruled out by induced GW overproduction violating the BBN bound. The gray shaded area shows the forecasted constraint from next generation CMB missions ($r<10^{-3}$) and the hatched area marks the strong-coupling region where $\lambda>1$ and quantum backreaction has to be taken into account.
We assume that oscillon formation becomes efficient at $M\lesssim0.05\Mpl$ \cite{Amin:2011hj}. We fixed $\beta=0.8$ in all cases, and for the generalized monodromy model with $q=5,10$ we conservatively assumed the same lifetime as for $q=0$. Varying these benchmark values shifts the boundary of the region ruled out by the BBN constraint, as we illustrate in \Cref{fig:parameter_space_variation_M_beta_tau}. For the $\alpha$-attractor T-model we explicitly vary $N_*$ to show the range of variation within the observational error bars, and analogous uncertainties are present for all models. To avoid further clutter in the plot, for the other models we fix the central value from $Planck$ for $n_s$ [and thus $N_*$ via \Cref{eq:n_s}] and focus on the variation with the model parameters $q$ and $p$. The ACT-preferred higher value for $n_s$ \cite{ACT:2025fju} would mainly yield a shift towards larger values of $N_*$, with effects similar to what is illustrated for the $\alpha$-attractor model.
}
\label{fig:parameter_space}
\end{figure*}
%
Even without immediate prospects for direct observation, we can use the constraint on the number of effective relativistic degrees of freedom, ${\Delta N_{\rm eff}<0.30}$ \cite{Planck:2018vyg}, derived from the CMB and BBN, to place bounds on the inflaton model parameters.
In terms of the energy density of gravitational waves, the BBN bound can be written as \cite{Caprini:2018mtu}
\begin{equation}
    \mathcal{A} \times \Omega_{\rm GW, res}^{\rm peak} \lesssim 0.068 \,,
    \label{eq:BBN_bound}
\end{equation}
where
\begin{equation}
    \mathcal{A} = \int_0^{k_{\rm osc}} d \ln k \left(\frac{k}{k_{\rm osc}}\right)^5 \Theta_{\rm uv}(k) \approx 0.23 \,.
\end{equation}
Using the expression \Cref{eq:Omega_GW_peak}, this bound can be used to place constraints on the model parameters, such as the inflaton mass $m$ and the self-interactions $g$ and $\lambda$.

The resulting parameter space in the $(m,g)$- and $(m,\lambda)$-planes is displayed in \Cref{fig:parameter_space}. The lower bound on $\lambda$ in \Cref{subfig:parameter_space_lam} [corresponding to the upper bound on $m$ in \Cref{subfig:parameter_space_g}] reflects the bound on $M$ stemming from the CMB constraint on $r$ listed in \Cref{tab:comparison} for some of the models. The lines converge around $m\sim 6\times 10^{-6}\Mpl$, as in this regime it is $M>\Mpl$ and the region of the potential probed by the CMB approaches the same quadratic minimum independent of $M$ for all models.
The boundary of the excluded region at lower $\lambda$ (higher $m$) marks the limit $M>0.05\Mpl$ below which oscillon formation becomes efficient \cite{Amin:2011hj} and the BBN constraint from GW overproduction becomes relevant.
Finally, the boundary at larger $\lambda$ (lower $m$) reflects the bound set by \Cref{eq:BBN_bound}.

In \Cref{fig:parameter_space_variation_M_beta_tau} we illustrate how the boundaries of the excluded region shift upon varying our benchmark values for the upper bound on $M$ for oscillon formation, the oscillon lifetime $\tau_{\rm osc}$, and their initial abundance $\beta$ (which are all model-dependent quantities). Also the coefficient $\alpha_m$, which is related to the oscillon mass via \Cref{eq:m_osc}, enters the GW amplitude and introduces an $\mathcal{O}(1)$ uncertainty. The resulting overall error bars emphasize the need for dedicated lattice simulations to measure the model parameters more precisely.

Observe how GW overproduction strongly constrains the viable parameter range for the combination of $m$ and $\lambda$, in particular. Importantly, for the monodromy potential of the form \Cref{eq:Monodromy-potential} with $q\gtrsim -5$ (including the logarithmic, plateau and axion monodromy models) we find that a significant region of parameter space where $\lambda<1$ is ruled out. Taking into account also the projected constraints from next-generation CMB experiments, we find that these models will be confined to the strong-coupling regime at $\lambda>1$. Consequently, if one demands $\lambda \lesssim 1$ in order to have a controlled, perturbative model, this class of potentials would then come into conflict with observations.
If one allows large $\lambda \gg 1$, the models may still be physically viable, but in this case strong quantum backreaction becomes important and needs to be taken into account when simulating the reheating phase after inflation. Allowing also larger negative values $q\lesssim-5$, allowed regions with $\lambda<1$ open up again.

Note also how the constraint in the $(m,g)$-plane is weaker than the one on $(m,\lambda)$, which follows from the relations \Cref{eq:couplings_self_IA}. Recalling from the expression \Cref{eq:Omega_GW_peak} that the BBN bound can be recast as a bound on $M$, the weaker dependence $g\propto M^{-1}$ compared to $\lambda\propto M^{-2}$ translates to a softer constraint.
In either case, our results demonstrate that induced GWs from oscillons provide a novel tool to probe and constrain inflationary models in a regime that is inaccessible by CMB observations alone.

On the one hand, our calculation may overestimate the induced GWs, as we are assuming an exactly monochromatic mass function and an instantaneous final decay. Although these assumptions are supported by results of simulations, the oscillons may form with some initial mass distribution, before they settle into the attractor solution at a fixed frequency, and in reality the decay takes a finite amount of time to complete. These effects may lead to a suppression of the induced GWs, as the decay will not happen exactly simultaneously everywhere and thus the transition from matter to radiation domination will be softened, see e.g.~\cite{Inomata:2019zqy, Pearce:2023kxp, Pearce:2025ywc} for studies investigating the effects of a gradual transition.

On the other hand, our constraints can be viewed as conservative, because we have not taken into account the enhanced GWs produced from the scale-invariant curvature power spectrum \cite{Inomata:2020lmk, Lozanov:2022yoy, Domenech:2024wao}, the ones induced during the matter dominated phase \cite{Papanikolaou:2020qtd}, or the GWs from preheating and oscillon formation \cite{Zhou:2013tsa, Antusch:2016con, Antusch:2017vga, Lozanov:2019ylm, Hiramatsu:2020obh}, which would all contribute to $N_{\rm eff}$ at BBN. Taking these into account as well would tighten the bounds we derived. Furthermore, in several of the models under consideration the oscillon lifetime is only bounded from below, meaning that the bounds will actually be stronger if the lifetime is significantly longer than the benchmarks in  \Cref{tab:comparison} we have assumed.

Finally, we have cut off the density contrast power spectrum \Cref{eq:Posc_Poiss} at the scale $k_{\rm osc}$, and in some cases the density contrast $\delta_{\rm osc}$ at the smallest scales near the peak at $k_{\rm osc}$ may become non-linear before the oscillons decay, as discussed in \Cref{app:k_NL}. However, GW production will likely continue in the non-linear regime and thus further contribute to $N_{\rm eff}$, see e.g.~\cite{Kawasaki:2023rfx, Fernandez:2023ddy}.

\section{Summary and conclusions}\label{sec:summary}
In this paper we have studied the induced GW spectrum generated due to the rapid decay of oscillons after they have dominated the post-inflationary Universe for several $e$-folds in an early matter-dominated phase. We have first discussed several general properties of oscillons and discussed three classes of potentials which sustain their production in \Cref{sec:dynamics_after_inflation}. \Cref{tab:comparison} summarizes key quantities for the models under consideration.
In \Cref{sec:perturbations_oscillons_domination} we discussed the spectrum of isocurvature perturbations which arise due to the Poisson noise associated with the discrete nature of the oscillons. The key result of this part is the analytical expression for the transfer function of the isocurvature induced curvature perturbation $\Phi_{\rm MD}$ in \Cref{eq:PhiIsoeMDInterpol}.
\Cref{sec:Induced_GWs} is devoted to the discussion of the induced GW spectrum. We presented an analytical expression for the resonant contribution to the spectrum in \Cref{eq:Omega_GW_res} and expressed the peak amplitude in terms of the oscillon parameters in \Cref{eq:Omega_GW_peak}. The spectrum today peaks at ultra-high frequencies at or above $\mathcal{O}(10^7)$Hz, making a direct detection challenging even for future experiments.
Importantly however, we found that the peak amplitude can surpass the BBN bound, yielding a novel way to constrain parameters of the inflaton potential in previously unexplored regions of parameter space. The resulting parameter space showing the allowed and ruled out regions is displayed in \Cref{fig:parameter_space}.
Our work demonstrates how novel constraints on inflationary models can be derived by using already existing data and accounting for the non-trivial dynamics during post-inflationary reheating.

As pointed out at the end of \Cref{sec:iGW_constraints}, our analysis could be extended by including other sources of GWs as well, which will also contribute to $N_{\rm eff}$ and thus tighten the constraints we have derived. The understanding of the non-linear regime at the smallest scales is another important direction for future work, which will most likely require numerical simulations.
Another relevant aspect to investigate in order to derive accurate constraints is the dependence of our constraints on experimental uncertainties, particularly given the new results of ACT, although we do not expect our main conclusions to be affected.
Key questions that need to be addressed with lattice simulations regard the precise details of oscillon formation, lifetime and decay. In particular, the upper bound on $M\ll \Mpl$, below which oscillon formation becomes efficient and which sets the boundary of the region ruled out by BBN, needs to be determined more precisely. Also the oscillon lifetime, for which in some models only lower bounds are available, needs to be measured in dedicated simulations. This is true also for the exact mass of individual oscillons, as well as the initial distribution of oscillon masses and the details of the final decay. We hope our results provide an incentive for these questions to be answered in future work.

Let us conclude by emphasizing that, given the precision of current and future CMB and GW data, a detailed understanding of the post-inflationary reheating dynamics is becoming a crucial aspect in determining details of the inflationary potential and constraining the microphysics that set the initial conditions for the hot Big Bang.

\begin{acknowledgments}
    We thank G.~Dom\`enech for useful discussions and M.~Gross for helpful correspondence.
    J.T.~is supported by the DFG under the Emmy-Noether program grant no.~DO 2574/1-1, project number 496592360. This work is supported in part by JSPS KAKENHI No.~24K00624.
\end{acknowledgments}

\appendix
\section{Relating potential parameters to CMB observables}\label{app:parameter-relations}
To illustrate the procedure of relating parameters of the inflationary potential to CMB observables, we will discuss the $\alpha$-attractor T-model, \Cref{eq:Tanh-potential} with $n=1$, in more detail. This potential has the advantage that analytical relations for all model parameters are available, while for other potentials some of these have to be obtained numerically.

Within the slow-roll approximation we can find relations between the model parameters $m$ and $M$, and the observables $A_s$, $n_s$ and $r$.
The Friedmann and Klein-Gordon equations read
\begin{equation}
    3 H^2 \Mpl^2 \approx V(\phi) \,, \quad 3 H \dot{\phi} + V_{\phi}(\phi) \approx 0 \,,
\end{equation}
where the dot denotes a derivative with respect to cosmic time, $\dot{}=d/dt$, and $V_{\phi}\equiv dV/d\phi$.

As usual, we define the end of inflation as the moment when the slow-roll condition is violated \cite{Baumann:2022mni}. In the slow-roll approximation, this corresponds to the potential slow-roll parameter $\epsilon_V$ becoming order unity, i.e.
\begin{equation}
    \epsilon_V(\phi_e) = \frac{\Mpl^2}{2}\left(\frac{V_{\phi}(\phi)}{V(\phi)}\right)^2\vline _{\phi_e} \simeq 1 \,.
    \label{eq:slow-roll-eps}
\end{equation}
Solving the above equation for the value of the field at the end of inflation, $\phi_e$, we find
\begin{align}
    \phi_e = \frac{M}{2} \text{arcsinh}\left(2\sqrt{2}\frac{\Mpl}{M} \right) \,,
\end{align}
and with this we obtain the Hubble parameter at the end of inflation
\begin{align}
    H_e \simeq \sqrt{\frac{V(\phi_e)}{3\Mpl^2}}
    \approx \frac{m M}{\sqrt{6}\Mpl} \,,
\end{align}
where the first equality is for a general slow-roll potential, and the second equality is for the $\alpha$-attractor model for which we took $M\ll \Mpl$, as required for oscillon formation.
We assume that oscillons form around the time, when inflation ends and the inflaton starts oscillating at the bottom of the potential, $H_{\rm f}\approx H_e$.

For the number of $e$-folds we find
\begin{align}
    N_* & = \int_{t_*}^{t_e} H dt \approx \int_{\phi_e}^{\phi_*} d\phi \, \frac{1}{\sqrt{2\epsilon_V}\Mpl} \label{eq:N_star}\\
    & \approx \frac{M^2}{4 \Mpl^2} \sinh^2\left(\frac{\phi_*}{M}\right)~\mbox{for $\alpha$-attractor},\nonumber
\end{align}
where $\phi_*$ is the value of the field when CMB scales exit the horizon. 
The second line demonstrates a characteristic feature of the $\alpha$-attractor T-model, namely that it allows for the analytical inversion of it to obtain an expression $\phi_*(N_*)$. 
Typically, such a relation has to be obtained numerically.

For the amplitude of the scalar power spectrum at the CMB pivot scale $k_*$ we obtain
\begin{align}
    A_s &= \frac{1}{8\pi^2 \epsilon_V(\phi_*)}\frac{H_*^2}{\Mpl^2} \label{eq:A_s} \\
    &    \approx \frac{N_*^2 m^2}{6 \pi^2\Mpl^2}~\mbox{for $\alpha$-attractor},\nonumber
\end{align}
which is to be evaluated at the time of horizon crossing.
The tensor-to-scalar ratio is given by
\begin{align}
    r &= 16 \epsilon_V(\phi_*)= \frac{2}{A_s \pi^2}\frac{H_*^2}{\Mpl^2} \\
   &\approx \frac{2 M^2}{N_*^2 \Mpl^2}~\mbox{for $\alpha$-attractor}\nonumber.
\end{align}

The last important inflationary observable is the scalar spectral index $n_s$, which can be computed as
\begin{align}
    n_s &= 1 - 6\epsilon_V(\phi_*) + 2\eta_V(\phi_*) \label{eq:n_s}\\
    &\approx 1-\frac{2}{N_*}~\mbox{for $\alpha$-attractor},\nonumber
\end{align}
where the second potential slow-roll parameter $\eta_V$ is defined as
\begin{equation}
    \eta_V = \Mpl^2 \frac{V_{\phi\phi}(\phi)}{V(\phi)} \,.
\end{equation}

\begin{figure}
\centering
\includegraphics[width=0.48\textwidth]{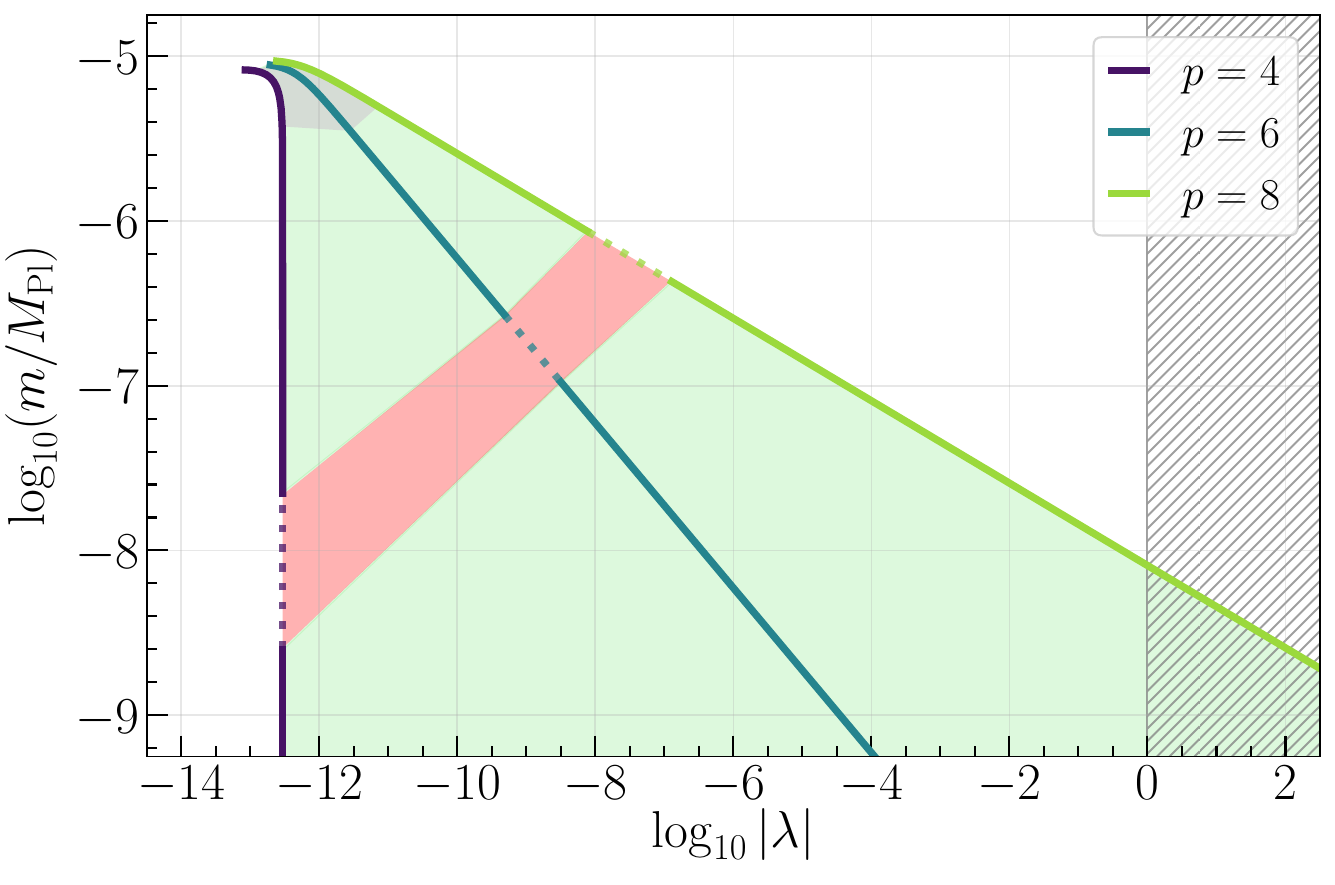}
\caption{The parameter space for the asymmetric hilltop potential \Cref{eq:Hilltop-potential} in the $(m,\lambda)$ plane with the same specifications as in \Cref{fig:parameter_space}. Note that $\lambda$ is negative and we plot the absolute value.
}
\label{fig:parameter_space_HT}
\end{figure}

A unique feature in the $\alpha$-attractor model is that both free parameters $m$ and $M$ can be fixed independently of each other in terms of the observable parameters $A_s$, $n_s$ and $r$.
The relations derived above coincide with the ones given in \cite{Lozanov:2019ylm}. For the logarithmic, axion monodromy and plateau potentials similar analytical relations can be obtained under the assumption $\left|\phi_*\right|\gg \left|\phi_e\right|$, and are provided in the appendix of \cite{Lozanov:2022yoy}.
\\
In the above we focused on the $\alpha$-attractor model, which allowed us to compute various quantities analytically. In the general case, particularly when $M\sim\Mpl$, such accurate analytical evaluations are not available for all the other potentials considered in this paper.
Therefore, for the bounds provided in \Cref{tab:comparison} and the parameter space plots \Cref{fig:parameter_space,fig:parameter_space_HT}, we compute the above relations numerically. The analytical approximations for $M\ll\Mpl$ serve as a validation of the numerical implementation.

Our general strategy for a precise numerical treatment is as follows. We first solve \Cref{eq:slow-roll-eps} for $\phi_e(M)$ and \Cref{eq:n_s} for $\phi_*(M)$. Then, we insert the obtained numerical function for $\phi_*(M)$ into the expression \Cref{eq:A_s} for $A_s$ and solve it for $m(M)$. For $n_s$ and $A_s$ we fix the $\textit{Planck}$ central values $n_s\approx 0.965$ and $A_s\approx 2.1 \times 10^{-9}$ \cite{Planck:2018vyg}. Finally, one can obtain $N_*(M)$ from \Cref{eq:N_star} by evaluating the integral and using the numerical results for $\phi_e(M)$ and $\phi_*(M)$.

\section{Evolution of isocurvature perturbations}\label{app:isocurvature_scenario}
To study the evolution of density perturbations and the generation of secondary gravitational waves we employ linear cosmological perturbation theory, see e.g.~\cite{Kodama:1984ziu, Malik:2008im, Baumann:2022mni} for more details. We consider a perturbed Friedmann–Lemaître–Robertson–Walker metric in the conformal Newtonian gauge taking the form
\begin{align}
    ds^2 = a^2(\tau) \big[& - (1-2 \Phi)\text{d}\tau^2 \nonumber \\
    &+ \left( (1 + 2 \Phi) \delta_{ij} + h_{ij}\right) \text{d}x^i \text{d}x^j \big] \,,
    \label{eq:FLRW}
\end{align}
where $a$ is the cosmic scale factor in terms of conformal time $\tau$, $\Phi$ denotes the gravitational potential, $h_{ij}$ are the transverse-traceless tensor perturbations, and we neglect vector perturbations and anisotropic stress.
The (background) energy densities of oscillons (``osc") and radiation (``r") evolve as
\begin{subequations} \label{eq:background_densities}
\begin{align}
    \rho_{\rm osc} &= 3 \beta \left(\frac{k_{\rm f}}{a_{\rm f}}\right)^2 \left(\frac{a}{a_{\rm f}}\right)^{-3} \,, \\
    \rho_{\rm r} &= 3 (1-\beta) \left(\frac{k_{\rm f}}{a_{\rm f}}\right)^2 \left(\frac{a}{a_{\rm f}}\right)^{-4} \,,
\end{align}
\end{subequations}
where $\beta\coloneq\rho_{\rm osc,f}/\rho_{\rm f}$ is the energy fraction that is converted into oscillons at the formation time denoted by the subscript ``f", and $k_{\rm f}$ is the scale that corresponds to the size of the (comoving) Hubble horizon at formation, $k_{\rm f}=a_{\rm f}H_{\rm f}$.
For convenience, we set $\Mpl=1$ in this Appendix.

\begin{figure}
\centering
\includegraphics[width=0.48\textwidth]{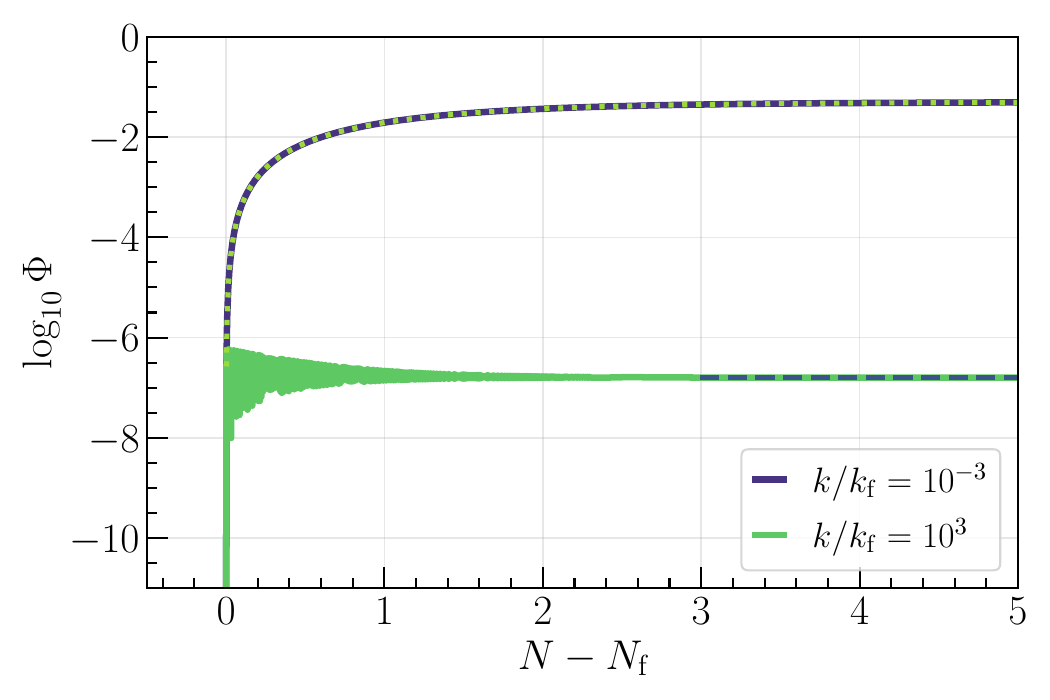}
\caption{Evolution of the gravitational potential $\Phi$ as a function of the number of $e$-folds since the time of oscillon formation, $N-N_{\rm f}$, with isocurvature initial conditions and for initial energy density fraction $\beta=0.8$. With solid lines we show the result of numerical integration of \Cref{eq:PhiEq,eq:SEq} for two modes which are initially sub- ($k/k_{\rm f}=10^{3}$) and superhorizon ($k/k_{\rm f}=10^{-3}$). The initial conditions are set at the time of formation, $\Phi_{\rm f}=0$, and we fixed $S_{\rm f}=1$ for the plot. The dotted and dashed lines superimposed on the numerical solution represent the analytical approximations \eqref{eq:Phi_chi_Super} and \eqref{eq:Phi_MD_Sub} for the super- and subhorizon regimes, respectively. See how the potential is quickly generated on both scales and approaches the plateau value fast.
}
\label{fig:Phi_iso_plot}
\end{figure}

The isocurvature (or entropy) perturbation is defined in terms of the density contrasts of the oscillon and radiation components as
\begin{equation}
    S = \delta_{\rm osc} - \frac{3}{4} \delta_{\rm r} \,,
\end{equation}
where $\delta_X = \delta\rho_X/\rho_X$ for $X\in\{\text{osc},\text{r}\}$.
The condition of initial isocurvature, $\Phi_{\rm f}=0$, implies
\begin{equation}
    \delta\rho_{\rm osc,f} + \delta\rho_{\rm r,f}=0 \,,
\end{equation}
which follows from the $00$-component of the perturbed Einstein equations, see e.g.~\cite{Malik:2008im} and \cite{Kodama:1986fg, Kodama:1986ud}.
Then one obtains for the initial amplitude of the isocurvature perturbation
\begin{equation}
    S_{\rm f} = \left(1+\frac{3}{4}\frac{\beta}{1-\beta}\right)\delta_{\rm osc,f} \,.
    \label{eq:S_f}
\end{equation}
From the perturbed Einstein equations and energy-momentum conservation, one can find a closed system of equations for $\Phi$ and $S$, given by
\begin{align}
    \Phi'' &+  3 \mathcal{H}(1+c_s^2)\Phi' + \left((1+3c_s^2)\mathcal{H}^2+2\mathcal{H}' +c_s^2 k^2\right)\Phi \nonumber \\
    &= \frac{1}{2}a^2c_s^2\rho_{\rm osc} S \,,
    \label{eq:PhiEq}
\end{align}
and
\begin{align}
    S'' &+ (1+(3 c_s^2-1))\mathcal{H}S' - \frac{1}{3}k^2(3 c_s^2-1)S \nonumber \\
    &= \frac{3}{2}\frac{c_s^2}{a^2\rho_{\rm r}}k^4 \Phi \,,
    \label{eq:SEq}
\end{align}
where $'\equiv d/d\tau$, $\calH\equiv a'/a$ denotes the comoving Hubble parameter, and the speed of sound is defined by
\begin{equation}
    c_s^2 = \, \frac{4}{3}\frac{\rho_{\rm r}}{3\rho_{\rm osc}+4\rho_{\rm r}} \,.
\end{equation}
In the superhorizon limit, $k\to 0$, the system can be solved analytically in terms of the scale factor $\chi\coloneq a/a_{\rm f}$, yielding
\begin{align}
    \Phi&(\chi;k\ll k_{\rm f}) = \frac{4 (\beta -1) S_{\rm f}}{15 (\beta -4) \beta ^3 \chi ^3}  \nonumber \\
    &\times \Bigg(\beta ^3 \left(3 \chi ^3+\chi ^2+4 \chi -8\right) \nonumber \\
    &\phantom{\times \Bigg(}-2 \beta ^2 \left(15 \sqrt{\beta  (\chi -1)+1}+3 \chi ^2+14 \chi -32\right) \nonumber \\ 
    &\phantom{\times \Bigg(}+8 \beta  \left(10 \sqrt{\beta  (\chi -1)+1}+3 \chi -13\right) \nonumber \\ 
    &\phantom{\times \Bigg(}-48 \left(\sqrt{\beta  (\chi -1)+1}-1\right)\Bigg) \,,
    \label{eq:Phi_chi_Super}
\end{align}
which grows rapidly from $\Phi(a_{\rm f})=0$ to the plateau value $\Phi_{\rm MD}=\Phi(\chi\gg 1)$ at late times. It is given by
\begin{equation} \label{eq:Phi_MD_Super}
    \Phi_{\rm MD}(k\ll k_{\rm f}) = \frac{4 (1-\beta)}{5 (4-\beta)} S_{\rm f} \,,
\end{equation}
which recovers the well known factor of $1/5$ in the limit $\beta \ll 1$ \cite{Kodama:1986fg, Kodama:1986ud}.
The isocurvature perturbation remains constant on superhorizon scales, ${S(\chi;k\ll k_{\rm f}) = S_{\rm f}}$.
Using \Cref{eq:background_densities} and the Friedmann equations, one can integrate \Cref{eq:PhiEq,eq:SEq} numerically also for small scales. We set the initial conditions at the formation time $a_{\rm in}=a_{\rm f}$ where $\Phi(a_{\rm in})=0$ and evolve them until some late time $a_{\rm end}\gg a_{\rm f}$, when $\Phi$ has reached the plateau value $\Phi_{\rm MD}$, which we can then read out.
We illustrate two examples for the evolution of $\Phi$ in \Cref{fig:Phi_iso_plot}.
%
%

By scanning over a range of values for $(\beta,k/k_{\rm f})$ and performing a fit to the obtained numerical data, we find that $\Phi_{\rm MD}$ scales as
\begin{align} \label{eq:Phi_MD_Sub}
    \Phi_{\rm MD}(k\gg k_{\rm f}) & \approx S_{\rm f} \times \mathcal{S}(\beta) \, \left(\frac{k}{k_{\rm f}}\right)^{-2} \,,
\end{align}
which is expected from the Poisson equation \labelcref{eq:Poisson_eq}.
We find that the suppression factor $\mathcal{S}(\beta)$ is well approximated by the fit
\begin{align} \label{eq:SuppressionFactor}
    \mathcal{S}(\beta) &\approx 1.48 (1-0.210\beta) \times  (1-\beta)\beta^2 \,,
\end{align}
in the regime $1/2\lesssim \beta<1$. \Cref{eq:SuppressionFactor} vanishes as $\beta\to1$ and recovers the $\propto \beta^2$ scaling, which is expected from known analytical approximations for $\beta\ll 1$ \cite{Kodama:1986fg, Kodama:1986ud}.
By interpolating between the large- and small-$k$ regimes of $\Phi_{\rm MD}$, \Cref{eq:Phi_MD_Super,eq:Phi_MD_Sub}, respectively, we obtain the result \Cref{eq:PhiIsoeMDInterpol} quoted in the main text.

The full $\beta$-dependence of the spectrum of $\Phi_{\rm MD}(k\gg k_{\rm f})$ is given by the product of \Cref{eq:SuppressionFactor} and the prefactor of \Cref{eq:S_f} and is collected in
\begin{equation} \label{eq:def_beta}
    \mathcal{C}(\beta)=\mathcal{S}(\beta)^2 \times \left(1+\frac{3}{4}\frac{\beta}{1-\beta}\right)^2 \,.
\end{equation}

\section{Non-linear scales}\label{app:k_NL}
During the oscillon-dominated period, density perturbations grow with the scale factor as ${\delta_{\rm osc} \propto a}$, indicating that for a long enough oscillon lifetime density fluctuations can become non-linear and thus violate the validity of linear perturbation theory. In order to estimate the scale, beyond which non-linearities can arise, we define $k_{\rm nl}$ such that at the time of oscillon decay, $t_{\rm rh}$, the density contrast reaches unity at that scale, $\delta_{\rm osc}(k_{\rm nl},t_{\rm rh})=1$ \cite{Inomata:2020lmk}.
We can then use the Poisson equation \Cref{eq:Poisson_eq} evaluated at the reheating time and at $k_{\rm nl}$ to find
\begin{equation}
    \Phi_{\rm MD}(k_{\rm nl}) = \frac{3}{2} \left(\frac{k_{\rm rh}}{k_{\rm nl}}\right)^2 \,.
\end{equation}
Then, we use the relations \Cref{eq:kosc_kf,eq:kf_krh} to express $k_{\rm nl}$ in term of $k_{\rm osc}$ and model parameters. As a rough estimate for the (generally scale-dependent) initial value of the isocurvature perturbation $S_{\rm f}$ we use \Cref{eq:S_f} and take $\delta_{\rm osc,f} \sim \mathcal{P}_{\delta_{\rm osc,f}}^{1/2}$ from \Cref{eq:Posc_Poiss}. Inserting typical values for the models under consideration, one can then numerically solve for $k_{\rm nl}/k_{\rm osc}$.
The results are rather sensitive to the precise numerical values and we cannot make a general statement. For the models with the longest oscillon lifetimes (e.g.~the axion monodromy potential), and for large values $\beta\sim 0.9$ and $M\sim 10^{-2}\Mpl$, the non-linear scale can be as low as $k_{\rm nl}\sim 10^{-2} k_{\rm osc}$, meaning that in these cases we are extrapolating linear perturbation theory into the non-linear regime at the smallest scales $\sim k_{\rm osc}$.
In contrast, for models which yield shorter lifetimes (like the hilltop model with $p=4$), and for smaller values of $\beta\sim 0.5$ and $M\lesssim 10^{-3}\Mpl$, we find $k_{\rm nl}\gtrsim k_{\rm osc}$, indicating that all scales which we are considering remain in the linear regime for these cases.

Let us mention also that, although $\delta_{\rm osc}$ may become order unity, the gravitational potential always remains small, $\Phi_{\rm MD}\ll 1$, due to the $k^2$ suppression in the Poisson equation. Therefore, one may argue that the use of linear perturbation theory to estimate the induced GWs sourced by $\Phi$ is justified, and the possible appearance of non-linearities in $\delta_{\rm osc}$ should not affect any of our main conclusions.

\section{Scalar induced gravitational waves}\label{app:SIGW}
%
\begin{figure}
\centering
\includegraphics[width=0.48\textwidth]{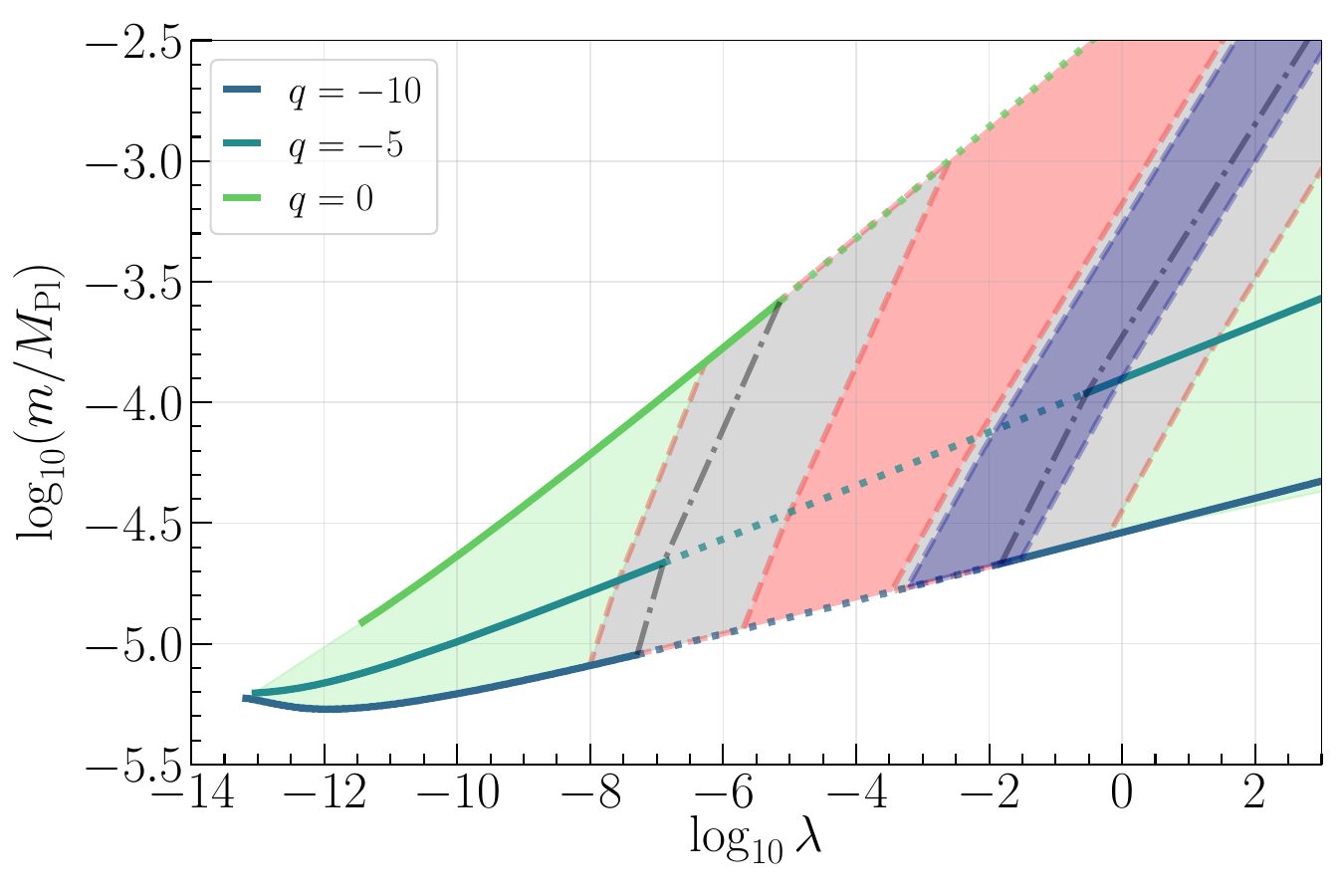}
\caption{Here we show the shift of the parameter space boundaries upon varying the threshold value for efficient oscillon formation between $0.01\le M_{\rm tr}\left[\Mpl\right]\le 0.1$ (gray shaded area, left), the initial abundance $0.5\le \beta \le 0.9$ (blue shaded), and the oscillon lifetime $10^{7}\le \tau_{\rm osc}\le 10^{9}$ (gray shaded, right). The benchmark values shown in \cref{subfig:parameter_space_lam} are marked with dash-dotted lines. For the sake of clarity of presentation we only illustrate three example models, but similar ranges of variation apply for the other models.
}
\label{fig:parameter_space_variation_M_beta_tau}
\end{figure}
%
In the computation of the scalar induced GWs we closely follow the formalism laid out in \cite{Domenech:2020ssp, Domenech:2021ztg, Domenech:2024wao} and we refer the interested reader to these references for more details.

The induced GW spectral energy density is given by the power spectrum of tensor modes as
\begin{equation} \label{eq:OmegaGWDefinition}
    \Omega_{\rm GW}(k) = \frac{k^2}{12 \mathcal{H}^2}\overline{\mathcal{P}_h(k,\tau)}\,,
\end{equation}
which should be evaluated at a time $\tau>\tau_{\rm rh}$ when the generation is completed and all tensor modes are deep inside the horizon and behave as free GWs.
The (oscillation-averaged) tensor spectrum $\overline{\mathcal{P}_h(k,\tau)}$ in turn is computed as a convolution of the tensor modes' Green's function with a source, which contains the quadratic scalar perturbations (i.e.~the gravitational potential $\Phi$ computed in \Cref{app:isocurvature_scenario}). The tensor power spectrum generated in the radiation era right after oscillon decay can then be written as \cite{Kohri:2018awv, Domenech:2020ssp}
\begin{align}
    &\overline{\mathcal{P}_{h,\rm RD}}(k,\tau;x\gg1) \approx \frac{c_s^4}{2048} \frac{x_{\rm rh}^8}{\bar{x}^2}  \int_{0}^{\infty} dv \int_{|1-v|}^{1+v} du \nonumber\\ &\left(4 v^2-\left(1+v^2-u^2\right)^2\right)^2 \overline{\mathcal{I}^2}(\bar{x},u,v) {\cal P}_{\Phi}(u k) {\cal P}_{\Phi}(v k) \,,
    \label{eq:TensorPowerSpectrum}
\end{align}
where $\bar{x}=x-x_{\rm rh}/2$ with $x=k\tau$, $x_{\rm rh}=2k/k_{\rm rh}$, and $c_s^2 = 1/3$ in radiation domination. For the scalar power spectrum ${\cal P}_{\Phi}$ we substitute \Cref{eq:PowerSpectrumPhi} in the GW kernel.
%
%
It is well-known that there is a resonant production of secondary GWs when the sum of the two momenta of the scalar modes matches that of the tensor mode, which happens at ${c_s(u+v)\simeq 1}$ \cite{Domenech:2021ztg}. For frequencies near this resonance one can analytically find the dominant contribution to the kernel $\overline{\mathcal{I}^2}$, which is given explicitly by \cite{Domenech:2020ssp, Domenech:2024wao}
\begin{align}
    \overline{\mathcal{I}^2_{\rm res}}(u,v) = \frac{1}{32} \text{Ci}\left(|(1-c_s(u+v))| x_{\rm rh}/2\right)^2 \,.
    \label{eq:OscIntegralRD2_res}
\end{align}
Inserting \Cref{eq:OscIntegralRD2_res} into \Cref{eq:TensorPowerSpectrum}, one can analytically evaluate the integrals to obtain the result given in \Cref{eq:Omega_GW_res}. The function $\Theta_{\rm uv}$, which appears due to momentum conservation and the cutoff we impose on the power spectrum \Cref{eq:Posc_Poiss} at $k_{\rm osc}$, is given by
\begin{align} \label{eq:Theta_UV}
    &\Theta_{\rm uv}(k) = \int_{-s_0}^{s_0} ds \frac{(1-s^2)^2}{(1-c_s^2 s^2)} \,,
\end{align}
with $s_0$ as defined in \cite{Inomata:2019ivs, Domenech:2020ssp,Domenech:2024wao}.

The dominant contribution to the low-frequency tail stems from large momenta $u \sim v\gg 1$ and is given by
\begin{align}
    \overline{\mathcal{I}^2_{\rm IR}}(u,v) = \frac{1}{8} \left(\text{Ci}(x_{\rm rh}/2)^2+\left(\pi/2 -\text{Si}(x_{\rm rh}/2)\right)^2\right) \,,
    \label{eq:OscIntegralRD2_LV}
\end{align}
with the sine and cosine integrals Si and Ci, respectively.

Once the tensor modes are deep inside the horizon, they behave as classical GWs and their energy density redshifts like that of radiation, $\rho_{\rm GW}\propto a^{-4}$. We can then directly relate the GW spectral density in the radiation era (denoted by RD) to the spectral energy density today as \cite{Inomata:2019ivs}
\begin{align}
    \Omega_{\rm GW,0}(f) h^2 \approx& 0.387 \left(\frac{g_*(T_{\rm RD})}{106.75}\right)^{-1/3} \Omega_{r,0}h^2 \nonumber \\
    &\times \Omega_{\rm GW, RD}(f)\,,\label{eq:Omega_GW_0}
\end{align}
where we used entropy conservation and $g_{*,s}(T_{\rm RD})=g_{*}(T_{\rm RD})$. The energy density fraction of radiation today is approximately $\Omega_{r,0}h^2 \approx 4.18\times 10^{-5}$ \cite{Planck:2018vyg}.

The spectral energy density of a stochastic GW background is related to the (dimensionless) characteristic strain $h_c$ by \cite{Aggarwal:2020olq}
\begin{equation}
    \Omega_{\rm GW}(f) = \frac{4\pi^2}{3 H_0^2} f^2 h_c^2(f) \,,
\end{equation}
which we use to transform the sensitivity curve of the resonant cavity experiment proposed in \cite{Herman:2022fau} for \Cref{fig:Omega_GW_0}.

\bibliography{references}

@article{Tomita:1967wkp,
    author = "Tomita, Kenji",
    title = "{Non-Linear Theory of Gravitational Instability in the Expanding Universe}",
    doi = "10.1143/PTP.37.831",
    journal = "Prog. Theor. Phys.",
    volume = "37",
    number = "5",
    pages = "831--846",
    year = "1967"
}

@article{Starobinsky:1980te,
    author = "Starobinsky, Alexei A.",
    editor = "Khalatnikov, I. M. and Mineev, V. P.",
    title = "{A New Type of Isotropic Cosmological Models Without Singularity}",
    doi = "10.1016/0370-2693(80)90670-X",
    journal = "Phys. Lett. B",
    volume = "91",
    pages = "99--102",
    year = "1980"
}

@article{Kazanas:1980tx,
    author = "Kazanas, D.",
    title = "{Dynamics of the Universe and Spontaneous Symmetry Breaking}",
    doi = "10.1086/183361",
    journal = "Astrophys. J. Lett.",
    volume = "241",
    pages = "L59--L63",
    year = "1980"
}

@article{Guth:1980zm,
    author = "Guth, Alan H.",
    editor = "Fang, Li-Zhi and Ruffini, R.",
    title = "{The Inflationary Universe: A Possible Solution to the Horizon and Flatness Problems}",
    reportNumber = "SLAC-PUB-2576",
    doi = "10.1103/PhysRevD.23.347",
    journal = "Phys. Rev. D",
    volume = "23",
    pages = "347--356",
    year = "1981"
}

@article{Mukhanov:1981xt,
    author = "Mukhanov, Viatcheslav F. and Chibisov, G. V.",
    title = "{Quantum Fluctuations and a Nonsingular Universe}",
    journal = "JETP Lett.",
    volume = "33",
    pages = "532--535",
    year = "1981"
}

@article{Linde:1981mu,
    author = "Linde, Andrei D.",
    editor = "Fang, Li-Zhi and Ruffini, R.",
    title = "{A New Inflationary Universe Scenario: A Possible Solution of the Horizon, Flatness, Homogeneity, Isotropy and Primordial Monopole Problems}",
    reportNumber = "LEBEDEV-81-229",
    doi = "10.1016/0370-2693(82)91219-9",
    journal = "Phys. Lett. B",
    volume = "108",
    pages = "389--393",
    year = "1982"
}

@article{Albrecht:1982wi,
    author = "Albrecht, Andreas and Steinhardt, Paul J.",
    editor = "Fang, Li-Zhi and Ruffini, R.",
    title = "{Cosmology for Grand Unified Theories with Radiatively Induced Symmetry Breaking}",
    reportNumber = "UPR-0185T",
    doi = "10.1103/PhysRevLett.48.1220",
    journal = "Phys. Rev. Lett.",
    volume = "48",
    pages = "1220--1223",
    year = "1982"
}

@article{Kodama:1984ziu,
    author = "Kodama, Hideo and Sasaki, Misao",
    title = "{Cosmological Perturbation Theory}",
    doi = "10.1143/PTPS.78.1",
    journal = "Prog. Theor. Phys. Suppl.",
    volume = "78",
    pages = "1--166",
    year = "1984"
}

@article{Sasaki:1986hm,
    author = "Sasaki, Misao",
    title = "{Large Scale Quantum Fluctuations in the Inflationary Universe}",
    reportNumber = "RRK-86-29",
    doi = "10.1143/PTP.76.1036",
    journal = "Prog. Theor. Phys.",
    volume = "76",
    pages = "1036",
    year = "1986"
}

@article{Kodama:1986fg,
    author = "Kodama, Hideo and Sasaki, Misao",
    title = "{Evolution of Isocurvature Perturbations. 1. Photon - Baryon Universe}",
    reportNumber = "UTAP-29",
    doi = "10.1142/S0217751X86000137",
    journal = "Int. J. Mod. Phys. A",
    volume = "1",
    pages = "265",
    year = "1986"
}

@article{Kodama:1986ud,
    author = "Kodama, Hideo and Sasaki, Misao",
    title = "{Evolution of Isocurvature Perturbations. 2. Radiation Dust Universe}",
    reportNumber = "UTAP-41",
    doi = "10.1142/S0217751X8700020X",
    journal = "Int. J. Mod. Phys. A",
    volume = "2",
    pages = "491",
    year = "1987"
}

@article{Mukhanov:1990me,
    author = "Mukhanov, Viatcheslav F. and Feldman, H. A. and Brandenberger, Robert H.",
    title = "{Theory of cosmological perturbations. Part 1. Classical perturbations. Part 2. Quantum theory of perturbations. Part 3. Extensions}",
    reportNumber = "BROWN-HET-796, BROWN-HET-800, BROWN-HET-780",
    doi = "10.1016/0370-1573(92)90044-Z",
    journal = "Phys. Rept.",
    volume = "215",
    pages = "203--333",
    year = "1992"
}

@article{Matarrese:1992rp,
  author       = {Matarrese, Sabino and Pantano, Ornella and Saez, Diego},
  title        = {{A General relativistic approach to the nonlinear evolution of collisionless matter}},
  reportnumber = {DFPD-92-A-39},
  doi          = {10.1103/PhysRevD.47.1311},
  journal      = {Phys. Rev. D},
  volume       = {47},
  pages        = {1311--1323},
  year         = {1993}
}

@article{Copeland:1995fq,
    author = "Copeland, Edmund J. and Gleiser, M. and Muller, H. -R.",
    title = "{Oscillons: Resonant configurations during bubble collapse}",
    eprint = "hep-ph/9503217",
    archivePrefix = "arXiv",
    reportNumber = "SUSX-TH-95-3-3, FERMILAB-PUB-95-021-A, DART-HEP-95-01",
    doi = "10.1103/PhysRevD.52.1920",
    journal = "Phys. Rev. D",
    volume = "52",
    pages = "1920--1933",
    year = "1995"
}

@article{Khlebnikov:1997di,
    author = "Khlebnikov, S. Y. and Tkachev, I. I.",
    title = "{Relic gravitational waves produced after preheating}",
    eprint = "hep-ph/9701423",
    archivePrefix = "arXiv",
    reportNumber = "PURD-TH-97-02, OSU-TA-01-97",
    doi = "10.1103/PhysRevD.56.653",
    journal = "Phys. Rev. D",
    volume = "56",
    pages = "653--660",
    year = "1997"
}

@book{Liddle:2000cg,
    author = "Liddle, Andrew R. and Lyth, David H.",
    title = "{Cosmological Inflation and Large-Scale Structure}",
    doi = "10.1017/CBO9781139175180",
    isbn = "978-0-521-57598-0, 978-0-521-82849-9",
    year = "2000",
    publisher={Cambridge University Press},
    place={Cambridge}
}

@article{Dymnikova:2000dy,
    author = "Dymnikova, I. and Koziel, L. and Khlopov, M. and Rubin, S.",
    title = "{Quasilumps from first order phase transitions}",
    eprint = "hep-th/0010120",
    archivePrefix = "arXiv",
    journal = "Grav. Cosmol.",
    volume = "6",
    pages = "311--318",
    year = "2000"
}

@article{Kasuya:2002zs,
    author = "Kasuya, S. and Kawasaki, M. and Takahashi, Fuminobu",
    title = "{I-balls}",
    eprint = "hep-ph/0209358",
    archivePrefix = "arXiv",
    doi = "10.1016/S0370-2693(03)00344-7",
    journal = "Phys. Lett. B",
    volume = "559",
    pages = "99--106",
    year = "2003"
}

@article{Boubekeur:2005zm,
    author = "Boubekeur, Lotfi and Lyth, David. H.",
    title = "{Hilltop inflation}",
    eprint = "hep-ph/0502047",
    archivePrefix = "arXiv",
    doi = "10.1088/1475-7516/2005/07/010",
    journal = "JCAP",
    volume = "07",
    pages = "010",
    year = "2005"
}

@article{Bassett:2005xm,
    author = "Bassett, Bruce A. and Tsujikawa, Shinji and Wands, David",
    title = "{Inflation dynamics and reheating}",
    eprint = "astro-ph/0507632",
    archivePrefix = "arXiv",
    doi = "10.1103/RevModPhys.78.537",
    journal = "Rev. Mod. Phys.",
    volume = "78",
    pages = "537--589",
    year = "2006"
}

@article{Easther:2006gt,
    author = "Easther, Richard and Lim, Eugene A.",
    title = "{Stochastic gravitational wave production after inflation}",
    eprint = "astro-ph/0601617",
    archivePrefix = "arXiv",
    doi = "10.1088/1475-7516/2006/04/010",
    journal = "JCAP",
    volume = "04",
    pages = "010",
    year = "2006"
}

@article{Ananda:2006af,
    author = "Ananda, Kishore N. and Clarkson, Chris and Wands, David",
    title = "{The Cosmological gravitational wave background from primordial density perturbations}",
    eprint = "gr-qc/0612013",
    archivePrefix = "arXiv",
    doi = "10.1103/PhysRevD.75.123518",
    journal = "Phys. Rev. D",
    volume = "75",
    pages = "123518",
    year = "2007"
}

@article{Baumann:2007zm,
    author = "Baumann, Daniel and Steinhardt, Paul J. and Takahashi, Keitaro and Ichiki, Kiyotomo",
    title = "{Gravitational Wave Spectrum Induced by Primordial Scalar Perturbations}",
    eprint = "hep-th/0703290",
    archivePrefix = "arXiv",
    doi = "10.1103/PhysRevD.76.084019",
    journal = "Phys. Rev. D",
    volume = "76",
    pages = "084019",
    year = "2007"
}

@article{Silverstein:2008sg,
    author = "Silverstein, Eva and Westphal, Alexander",
    title = "{Monodromy in the CMB: Gravity Waves and String Inflation}",
    eprint = "0803.3085",
    archivePrefix = "arXiv",
    primaryClass = "hep-th",
    reportNumber = "SU-ITP-08-07, SLAC-PUB-13183",
    doi = "10.1103/PhysRevD.78.106003",
    journal = "Phys. Rev. D",
    volume = "78",
    pages = "106003",
    year = "2008"
}

@article{Gleiser:2008ty,
    author = "Gleiser, Marcelo and Sicilia, David",
    title = "{Analytical Characterization of Oscillon Energy and Lifetime}",
    eprint = "0804.0791",
    archivePrefix = "arXiv",
    primaryClass = "hep-th",
    doi = "10.1103/PhysRevLett.101.011602",
    journal = "Phys. Rev. Lett.",
    volume = "101",
    pages = "011602",
    year = "2008"
}

@article{McAllister:2008hb,
    author = "McAllister, Liam and Silverstein, Eva and Westphal, Alexander",
    title = "{Gravity Waves and Linear Inflation from Axion Monodromy}",
    eprint = "0808.0706",
    archivePrefix = "arXiv",
    primaryClass = "hep-th",
    reportNumber = "SLAC-PUB-13357, SU-ITP-08-15",
    doi = "10.1103/PhysRevD.82.046003",
    journal = "Phys. Rev. D",
    volume = "82",
    pages = "046003",
    year = "2010"
}

@article{Malik:2008im,
    author = "Malik, Karim A. and Wands, David",
    title = "{Cosmological perturbations}",
    eprint = "0809.4944",
    archivePrefix = "arXiv",
    primaryClass = "astro-ph",
    doi = "10.1016/j.physrep.2009.03.001",
    journal = "Phys. Rept.",
    volume = "475",
    pages = "1--51",
    year = "2009"
}

@inproceedings{Baumann:2009ds,
    author = "Baumann, Daniel",
    title = "{Inflation}",
    booktitle = "{Theoretical Advanced Study Institute in Elementary Particle Physics}: {Physics of the Large and the Small}",
    eprint = "0907.5424",
    archivePrefix = "arXiv",
    primaryClass = "hep-th",
    reportNumber = "TASI-2009",
    doi = "10.1142/9789814327183_0010",
    pages = "523--686",
    year = "2011"
}

@article{Amin:2010jq,
    author = "Amin, Mustafa A. and Shirokoff, David",
    title = "{Flat-top oscillons in an expanding universe}",
    eprint = "1002.3380",
    archivePrefix = "arXiv",
    primaryClass = "astro-ph.CO",
    doi = "10.1103/PhysRevD.81.085045",
    journal = "Phys. Rev. D",
    volume = "81",
    pages = "085045",
    year = "2010"
}

@article{Hertzberg:2010yz,
    author = "Hertzberg, Mark P.",
    title = "{Quantum Radiation of Oscillons}",
    eprint = "1003.3459",
    archivePrefix = "arXiv",
    primaryClass = "hep-th",
    reportNumber = "MIT-CTP-4130",
    doi = "10.1103/PhysRevD.82.045022",
    journal = "Phys. Rev. D",
    volume = "82",
    pages = "045022",
    year = "2010"
}

@article{Amin:2010xe,
    author = "Amin, Mustafa A.",
    title = "{Inflaton fragmentation: Emergence of pseudo-stable inflaton lumps (oscillons) after inflation}",
    eprint = "1006.3075",
    archivePrefix = "arXiv",
    primaryClass = "astro-ph.CO",
    month = "6",
    year = "2010",
    journal=""
}

@article{Amin:2010dc,
    author = "Amin, Mustafa A. and Easther, Richard and Finkel, Hal",
    title = "{Inflaton Fragmentation and Oscillon Formation in Three Dimensions}",
    eprint = "1009.2505",
    archivePrefix = "arXiv",
    primaryClass = "astro-ph.CO",
    doi = "10.1088/1475-7516/2010/12/001",
    journal = "JCAP",
    volume = "12",
    pages = "001",
    year = "2010"
}

@article{Amin:2011hj,
    author = "Amin, Mustafa A. and Easther, Richard and Finkel, Hal and Flauger, Raphael and Hertzberg, Mark P.",
    title = "{Oscillons After Inflation}",
    eprint = "1106.3335",
    archivePrefix = "arXiv",
    primaryClass = "astro-ph.CO",
    doi = "10.1103/PhysRevLett.108.241302",
    journal = "Phys. Rev. Lett.",
    volume = "108",
    pages = "241302",
    year = "2012"
}

@article{Salmi:2012ta,
    author = "Salmi, Petja and Hindmarsh, Mark",
    title = "{Radiation and Relaxation of Oscillons}",
    eprint = "1201.1934",
    archivePrefix = "arXiv",
    primaryClass = "hep-th",
    doi = "10.1103/PhysRevD.85.085033",
    journal = "Phys. Rev. D",
    volume = "85",
    pages = "085033",
    year = "2012"
}

@article{Martin:2013tda,
    author = "Martin, Jerome and Ringeval, Christophe and Vennin, Vincent",
    title = "{Encyclop{\ae}dia Inflationaris}: {Opiparous Edition}",
    eprint = "1303.3787",
    archivePrefix = "arXiv",
    primaryClass = "astro-ph.CO",
    doi = "10.1016/j.dark.2024.101653",
    journal = "Phys. Dark Univ.",
    volume = "5-6",
    pages = "75--235",
    year = "2014"
}

@article{Zhou:2013tsa,
    author = "Zhou, Shuang-Yong and Copeland, Edmund J. and Easther, Richard and Finkel, Hal and Mou, Zong-Gang and Saffin, Paul M.",
    title = "{Gravitational Waves from Oscillon Preheating}",
    eprint = "1304.6094",
    archivePrefix = "arXiv",
    primaryClass = "astro-ph.CO",
    doi = "10.1007/JHEP10(2013)026",
    journal = "JHEP",
    volume = "10",
    pages = "026",
    year = "2013"
}

@article{Kallosh:2013hoa,
    author = "Kallosh, Renata and Linde, Andrei",
    title = "{Universality Class in Conformal Inflation}",
    eprint = "1306.5220",
    archivePrefix = "arXiv",
    primaryClass = "hep-th",
    doi = "10.1088/1475-7516/2013/07/002",
    journal = "JCAP",
    volume = "07",
    pages = "002",
    year = "2013"
}

@article{Kallosh:2013yoa,
    author = "Kallosh, Renata and Linde, Andrei and Roest, Diederik",
    title = "{Superconformal Inflationary $\alpha$-Attractors}",
    eprint = "1311.0472",
    archivePrefix = "arXiv",
    primaryClass = "hep-th",
    doi = "10.1007/JHEP11(2013)198",
    journal = "JHEP",
    volume = "11",
    pages = "198",
    year = "2013"
}

@article{Saffin:2014yka,
    author = "Saffin, Paul M. and Tognarelli, Paul and Tranberg, Anders",
    title = "{Oscillon Lifetime in the Presence of Quantum Fluctuations}",
    eprint = "1401.6168",
    archivePrefix = "arXiv",
    primaryClass = "hep-ph",
    doi = "10.1007/JHEP08(2014)125",
    journal = "JHEP",
    volume = "08",
    pages = "125",
    year = "2014"
}

@article{McAllister:2014mpa,
    author = "McAllister, Liam and Silverstein, Eva and Westphal, Alexander and Wrase, Timm",
    title = "{The Powers of Monodromy}",
    eprint = "1405.3652",
    archivePrefix = "arXiv",
    primaryClass = "hep-th",
    reportNumber = "SU/ITP-14/13, SLAC-PUB-15962, DESY-14-078, SU-ITP-14-13",
    doi = "10.1007/JHEP09(2014)123",
    journal = "JHEP",
    volume = "09",
    pages = "123",
    year = "2014"
}

@article{Takeda:2014qma,
    author = "Takeda, Naoyuki and Watanabe, Yuki",
    title = "{No quasistable scalaron lump forms after $R^2$ inflation}",
    eprint = "1405.3830",
    archivePrefix = "arXiv",
    primaryClass = "astro-ph.CO",
    reportNumber = "ICRR-REPORT-681-2014-7, RESCEU-10-14",
    doi = "10.1103/PhysRevD.90.023519",
    journal = "Phys. Rev. D",
    volume = "90",
    number = "2",
    pages = "023519",
    year = "2014"
}

@article{Kawasaki:2015vga,
    author = "Kawasaki, Masahiro and Takahashi, Fuminobu and Takeda, Naoyuki",
    title = "{Adiabatic Invariance of Oscillons/I-balls}",
    eprint = "1508.01028",
    archivePrefix = "arXiv",
    primaryClass = "hep-th",
    reportNumber = "TU-1004, IPMU-15-0128",
    doi = "10.1103/PhysRevD.92.105024",
    journal = "Phys. Rev. D",
    volume = "92",
    number = "10",
    pages = "105024",
    year = "2015"
}

@article{Antusch:2015ziz,
    author = "Antusch, Stefan and Orani, Stefano",
    title = "{Impact of other scalar fields on oscillons after hilltop inflation}",
    eprint = "1511.02336",
    archivePrefix = "arXiv",
    primaryClass = "hep-ph",
    doi = "10.1088/1475-7516/2016/03/026",
    journal = "JCAP",
    volume = "03",
    pages = "026",
    year = "2016"
}

@article{Antusch:2016con,
    author = "Antusch, Stefan and Cefala, Francesco and Orani, Stefano",
    title = "{Gravitational waves from oscillons after inflation}",
    eprint = "1607.01314",
    archivePrefix = "arXiv",
    primaryClass = "astro-ph.CO",
    doi = "10.1103/PhysRevLett.118.011303",
    journal = "Phys. Rev. Lett.",
    volume = "118",
    number = "1",
    pages = "011303",
    year = "2017",
    note = "[Erratum: Phys.Rev.Lett. 120, 219901 (2018)]"
}

@article{Lozanov:2016hid,
    author = "Lozanov, Kaloian D. and Amin, Mustafa A.",
    title = "{Equation of State and Duration to Radiation Domination after Inflation}",
    eprint = "1608.01213",
    archivePrefix = "arXiv",
    primaryClass = "astro-ph.CO",
    doi = "10.1103/PhysRevLett.119.061301",
    journal = "Phys. Rev. Lett.",
    volume = "119",
    number = "6",
    pages = "061301",
    year = "2017"
}

@article{Kim:2017duj,
    author = "Kim, Jinsu and McDonald, John",
    title = "{Inflaton Condensate Fragmentation: Analytical Conditions and Application to $\alpha$-Attractor Models}",
    eprint = "1702.08777",
    archivePrefix = "arXiv",
    primaryClass = "astro-ph.CO",
    doi = "10.1103/PhysRevD.95.123537",
    journal = "Phys. Rev. D",
    volume = "95",
    number = "12",
    pages = "123537",
    year = "2017"
}

@article{Lozanov:2017hjm,
    author = "Lozanov, Kaloian D. and Amin, Mustafa A.",
    title = "{Self-resonance after inflation: oscillons, transients and radiation domination}",
    eprint = "1710.06851",
    archivePrefix = "arXiv",
    primaryClass = "astro-ph.CO",
    doi = "10.1103/PhysRevD.97.023533",
    journal = "Phys. Rev. D",
    volume = "97",
    number = "2",
    pages = "023533",
    year = "2018"
}

@article{Antusch:2017vga,
    author = "Antusch, Stefan and Cefala, Francesco and Orani, Stefano",
    title = "{What can we learn from the stochastic gravitational wave background produced by oscillons?}",
    eprint = "1712.03231",
    archivePrefix = "arXiv",
    primaryClass = "astro-ph.CO",
    doi = "10.1088/1475-7516/2018/03/032",
    journal = "JCAP",
    volume = "03",
    pages = "032",
    year = "2018"
}

@article{Cotner:2018vug,
    author = "Cotner, Eric and Kusenko, Alexander and Takhistov, Volodymyr",
    title = "{Primordial Black Holes from Inflaton Fragmentation into Oscillons}",
    eprint = "1801.03321",
    archivePrefix = "arXiv",
    primaryClass = "astro-ph.CO",
    reportNumber = "IPMU18-0008",
    doi = "10.1103/PhysRevD.98.083513",
    journal = "Phys. Rev. D",
    volume = "98",
    number = "8",
    pages = "083513",
    year = "2018"
}

@article{Kohri:2018awv,
    author = "Kohri, Kazunori and Terada, Takahiro",
    title = "{Semianalytic calculation of gravitational wave spectrum nonlinearly induced from primordial curvature perturbations}",
    eprint = "1804.08577",
    archivePrefix = "arXiv",
    primaryClass = "gr-qc",
    reportNumber = "KEK-TH-2046, KEK-COSMO-223",
    doi = "10.1103/PhysRevD.97.123532",
    journal = "Phys. Rev. D",
    volume = "97",
    number = "12",
    pages = "123532",
    year = "2018"
}

@article{Planck:2018vyg,
    author = "Aghanim, N. and others",
    collaboration = "Planck",
    title = "{Planck 2018 results. VI. Cosmological parameters}",
    eprint = "1807.06209",
    archivePrefix = "arXiv",
    primaryClass = "astro-ph.CO",
    doi = "10.1051/0004-6361/201833910",
    journal = "Astron. Astrophys.",
    volume = "641",
    pages = "A6",
    year = "2020",
    note = "[Erratum: Astron.Astrophys. 652, C4 (2021)]"
}

@article{Planck:2018jri,
    author = "Akrami, Y. and others",
    collaboration = "Planck",
    title = "{Planck 2018 results. X. Constraints on inflation}",
    eprint = "1807.06211",
    archivePrefix = "arXiv",
    primaryClass = "astro-ph.CO",
    doi = "10.1051/0004-6361/201833887",
    journal = "Astron. Astrophys.",
    volume = "641",
    pages = "A10",
    year = "2020"
}

@article{SimonsObservatory:2018koc,
    author = "Ade, Peter and others",
    collaboration = "Simons Observatory",
    title = "{The Simons Observatory: Science goals and forecasts}",
    eprint = "1808.07445",
    archivePrefix = "arXiv",
    primaryClass = "astro-ph.CO",
    doi = "10.1088/1475-7516/2019/02/056",
    journal = "JCAP",
    volume = "02",
    pages = "056",
    year = "2019"
}

@article{Caprini:2018mtu,
    author = "Caprini, Chiara and Figueroa, Daniel G.",
    title = "{Cosmological Backgrounds of Gravitational Waves}",
    eprint = "1801.04268",
    archivePrefix = "arXiv",
    primaryClass = "astro-ph.CO",
    doi = "10.1088/1361-6382/aac608",
    journal = "Class. Quant. Grav.",
    volume = "35",
    number = "16",
    pages = "163001",
    year = "2018"
}

@article{Ibe:2019vyo,
    author = "Ibe, Masahiro and Kawasaki, Masahiro and Nakano, Wakutaka and Sonomoto, Eisuke",
    title = "{Decay of I-ball/Oscillon in Classical Field Theory}",
    eprint = "1901.06130",
    archivePrefix = "arXiv",
    primaryClass = "hep-ph",
    reportNumber = "IPMU19-0002",
    doi = "10.1007/JHEP04(2019)030",
    journal = "JHEP",
    volume = "04",
    pages = "030",
    year = "2019"
}

@article{Lozanov:2019ylm,
    author = "Lozanov, Kaloian D. and Amin, Mustafa A.",
    title = "{Gravitational perturbations from oscillons and transients after inflation}",
    eprint = "1902.06736",
    archivePrefix = "arXiv",
    primaryClass = "astro-ph.CO",
    doi = "10.1103/PhysRevD.99.123504",
    journal = "Phys. Rev. D",
    volume = "99",
    number = "12",
    pages = "123504",
    year = "2019"
}

@article{Amin:2019ums,
    author = "Amin, Mustafa A. and Mocz, Philip",
    title = "{Formation, gravitational clustering, and interactions of nonrelativistic solitons in an expanding universe}",
    eprint = "1902.07261",
    archivePrefix = "arXiv",
    primaryClass = "astro-ph.CO",
    doi = "10.1103/PhysRevD.100.063507",
    journal = "Phys. Rev. D",
    volume = "100",
    number = "6",
    pages = "063507",
    year = "2019"
}

@article{Inomata:2019ivs,
    author = "Inomata, Keisuke and Kohri, Kazunori and Nakama, Tomohiro and Terada, Takahiro",
    title = "{Enhancement of Gravitational Waves Induced by Scalar Perturbations due to a Sudden Transition from an Early Matter Era to the Radiation Era}",
    eprint = "1904.12879",
    archivePrefix = "arXiv",
    primaryClass = "astro-ph.CO",
    reportNumber = "IPMU 19-0067, KEK-TH-2122, KEK-Cosmo-237",
    doi = "10.1103/PhysRevD.108.049901",
    journal = "Phys. Rev. D",
    volume = "100",
    pages = "043532",
    year = "2019",
    note = "[Erratum: Phys.Rev.D 108, 049901 (2023)]"
}

@article{Inomata:2019zqy,
    author = "Inomata, Keisuke and Kohri, Kazunori and Nakama, Tomohiro and Terada, Takahiro",
    title = "{Gravitational Waves Induced by Scalar Perturbations during a Gradual Transition from an Early Matter Era to the Radiation Era}",
    eprint = "1904.12878",
    archivePrefix = "arXiv",
    primaryClass = "astro-ph.CO",
    reportNumber = "IPMU 19-0066, KEK-TH-2121, KEK-Cosmo-236",
    doi = "10.1088/1475-7516/2019/10/071",
    journal = "JCAP",
    volume = "10",
    pages = "071",
    year = "2019",
    note = "[Erratum: JCAP 08, E01 (2023)]"
}

@article{Olle:2019kbo,
    author = "Oll{\'e}, Jan and Pujol{\`a}s, Oriol and Rompineve, Fabrizio",
    title = "{Oscillons and Dark Matter}",
    eprint = "1906.06352",
    archivePrefix = "arXiv",
    primaryClass = "hep-ph",
    doi = "10.1088/1475-7516/2020/02/006",
    journal = "JCAP",
    volume = "02",
    pages = "006",
    year = "2020"
}

@article{Gleiser:2019rvw,
    author = "Gleiser, Marcelo and Krackow, Max",
    title = "{Resonant configurations in scalar field theories: Can some oscillons live forever?}",
    eprint = "1906.04070",
    archivePrefix = "arXiv",
    primaryClass = "hep-th",
    doi = "10.1103/PhysRevD.100.116005",
    journal = "Phys. Rev. D",
    volume = "100",
    number = "11",
    pages = "116005",
    year = "2019"
}

@article{Cotner:2019ykd,
    author = "Cotner, Eric and Kusenko, Alexander and Sasaki, Misao and Takhistov, Volodymyr",
    title = "{Analytic Description of Primordial Black Hole Formation from Scalar Field Fragmentation}",
    eprint = "1907.10613",
    archivePrefix = "arXiv",
    primaryClass = "astro-ph.CO",
    reportNumber = "IPMU19-0063, YITP-19-31",
    doi = "10.1088/1475-7516/2019/10/077",
    journal = "JCAP",
    volume = "10",
    pages = "077",
    year = "2019"
}

@article{Lozanov:2019jxc,
    author = "Lozanov, Kaloian D.",
    title = "{Lectures on Reheating after Inflation}",
    eprint = "1907.04402",
    archivePrefix = "arXiv",
    primaryClass = "astro-ph.CO",
    month = "7",
    year = "2019",
    journal=""
}

@article{Antusch:2019qrr,
    author = "Antusch, Stefan and Cefal\`a, Francesco and Torrent\'\i{}, Francisco",
    title = "{Properties of Oscillons in Hilltop Potentials: energies, shapes, and lifetimes}",
    eprint = "1907.00611",
    archivePrefix = "arXiv",
    primaryClass = "hep-ph",
    doi = "10.1088/1475-7516/2019/10/002",
    journal = "JCAP",
    volume = "10",
    pages = "002",
    year = "2019"
}

@article{Inomata:2020lmk,
    author = "Inomata, Keisuke and Kawasaki, Masahiro and Mukaida, Kyohei and Terada, Takahiro and Yanagida, Tsutomu T.",
    title = "{Gravitational Wave Production right after a Primordial Black Hole Evaporation}",
    eprint = "2003.10455",
    archivePrefix = "arXiv",
    primaryClass = "astro-ph.CO",
    reportNumber = "IPMU 20-0029, DESY 20-042, DESY-20-042, CTPU-PTC-20-05",
    doi = "10.1103/PhysRevD.101.123533",
    journal = "Phys. Rev. D",
    volume = "101",
    number = "12",
    pages = "123533",
    year = "2020"
}

@article{Zhang:2020bec,
    author = "Zhang, Hong-Yi and Amin, Mustafa A. and Copeland, Edmund J. and Saffin, Paul M. and Lozanov, Kaloian D.",
    title = "{Classical Decay Rates of Oscillons}",
    eprint = "2004.01202",
    archivePrefix = "arXiv",
    primaryClass = "hep-th",
    doi = "10.1088/1475-7516/2020/07/055",
    journal = "JCAP",
    volume = "07",
    pages = "055",
    year = "2020"
}

@article{Allahverdi:2020bys,
    author = "Allahverdi, Rouzbeh and others",
    title = "{The First Three Seconds: a Review of Possible Expansion Histories of the Early Universe}",
    eprint = "2006.16182",
    archivePrefix = "arXiv",
    primaryClass = "astro-ph.CO",
    reportNumber = "FERMILAB-PUB-20-242-A, KCL-PH-TH/2020-33, KEK-Cosmo-257,
  KEK-TH-2231, IPMU20-0070, PI/UAN-2020-674FT, RUP-20-22",
    doi = "10.21105/astro.2006.16182",
    journal = "Open J. Astrophys.",
    volume = "4",
    pages = "astro.2006.16182",
    year = "2021"
}

@article{Papanikolaou:2020qtd,
    author = "Papanikolaou, Theodoros and Vennin, Vincent and Langlois, David",
    title = "{Gravitational waves from a universe filled with primordial black holes}",
    eprint = "2010.11573",
    archivePrefix = "arXiv",
    primaryClass = "astro-ph.CO",
    doi = "10.1088/1475-7516/2021/03/053",
    journal = "JCAP",
    volume = "03",
    pages = "053",
    year = "2021"
}

@article{Hiramatsu:2020obh,
    author = "Hiramatsu, Takashi and Sfakianakis, Evangelos I. and Yamaguchi, Masahide",
    title = "{Gravitational wave spectra from oscillon formation after inflation}",
    eprint = "2011.12201",
    archivePrefix = "arXiv",
    primaryClass = "hep-ph",
    reportNumber = "Nikhef 2020-028, RUP-20-33",
    doi = "10.1007/JHEP03(2021)021",
    journal = "JHEP",
    volume = "03",
    pages = "021",
    year = "2021"
}

@article{Aggarwal:2020olq,
    author = "Aggarwal, Nancy and others",
    title = "{Challenges and opportunities of gravitational-wave searches at MHz to GHz frequencies}",
    eprint = "2011.12414",
    archivePrefix = "arXiv",
    primaryClass = "gr-qc",
    reportNumber = "CERN-TH-2020-185, HIP-2020-28/TH, DESY 20-195, CERN-TH-2020-185, HIP-2020-28/TH, DESY 20-195",
    doi = "10.1007/s41114-021-00032-5",
    journal = "Living Rev. Rel.",
    volume = "24",
    number = "1",
    pages = "4",
    year = "2021"
}

@article{Domenech:2020ssp,
    author = "Dom{\`e}nech, Guillem and Lin, Chunshan and Sasaki, Misao",
    title = "{Gravitational wave constraints on the primordial black hole dominated early universe}",
    eprint = "2012.08151",
    archivePrefix = "arXiv",
    primaryClass = "gr-qc",
    reportNumber = "YITP-20-156",
    doi = "10.1088/1475-7516/2021/11/E01",
    journal = "JCAP",
    volume = "04",
    pages = "062",
    year = "2021",
    note = "[Erratum: JCAP 11, E01 (2021)]"
}

@article{Domenech:2021wkk,
    author = "Dom{\`e}nech, Guillem and Takhistov, Volodymyr and Sasaki, Misao",
    title = "{Exploring evaporating primordial black holes with gravitational waves}",
    eprint = "2105.06816",
    archivePrefix = "arXiv",
    primaryClass = "astro-ph.CO",
    reportNumber = "IPMU21-0028, YITP-21-44",
    doi = "10.1016/j.physletb.2021.136722",
    journal = "Phys. Lett. B",
    volume = "823",
    pages = "136722",
    year = "2021"
}

@article{Domenech:2021ztg,
    author = "Dom\`enech, Guillem",
    title = "{Scalar Induced Gravitational Waves Review}",
    eprint = "2109.01398",
    archivePrefix = "arXiv",
    primaryClass = "gr-qc",
    doi = "10.3390/universe7110398",
    journal = "Universe",
    volume = "7",
    number = "11",
    pages = "398",
    year = "2021"
}

@article{Imagawa:2021sxt,
    author = "Imagawa, Kaname and Kawasaki, Masahiro and Murai, Kai and Nakatsuka, Hiromasa and Sonomoto, Eisuke",
    title = "{Free streaming length of axion-like particle after oscillon/I-ball decays}",
    eprint = "2110.05790",
    archivePrefix = "arXiv",
    primaryClass = "hep-ph",
    doi = "10.1088/1475-7516/2023/02/024",
    journal = "JCAP",
    volume = "02",
    pages = "024",
    year = "2023"
}

@article{Kim:2021ipz,
    author = "Kim, Jinsu and McDonald, John",
    title = "{General analytical conditions for inflaton fragmentation: Quick and easy tests for its occurrence}",
    eprint = "2111.12474",
    archivePrefix = "arXiv",
    primaryClass = "astro-ph.CO",
    reportNumber = "CERN-TH-2021-202",
    doi = "10.1103/PhysRevD.105.063508",
    journal = "Phys. Rev. D",
    volume = "105",
    number = "6",
    pages = "063508",
    year = "2022"
}

@article{Berlin:2021txa,
    author = {Berlin, Asher and Blas, Diego and Tito D'Agnolo, Raffaele and Ellis, Sebastian A. R. and Harnik, Roni and Kahn, Yonatan and Sch{\"u}tte-Engel, Jan},
    title = "{Detecting high-frequency gravitational waves with microwave cavities}",
    eprint = "2112.11465",
    archivePrefix = "arXiv",
    primaryClass = "hep-ph",
    reportNumber = "FERMILAB-PUB-21-724-SQMS-T",
    doi = "10.1103/PhysRevD.105.116011",
    journal = "Phys. Rev. D",
    volume = "105",
    number = "11",
    pages = "116011",
    year = "2022"
}

@book{Baumann:2022mni,
    author = "Baumann, Daniel",
    title = "{Cosmology}",
    doi = "10.1017/9781108937092",
    isbn = "978-1-108-93709-2, 978-1-108-83807-8",
    publisher = "Cambridge University Press",
    month = "7",
    year = "2022"
}

@article{Domcke:2022rgu,
    author = "Domcke, Valerie and Garcia-Cely, Camilo and Rodd, Nicholas L.",
    title = "{Novel Search for High-Frequency Gravitational Waves with Low-Mass Axion Haloscopes}",
    eprint = "2202.00695",
    archivePrefix = "arXiv",
    primaryClass = "hep-ph",
    reportNumber = "DESY-22-017, CERN-TH-2022-010",
    doi = "10.1103/PhysRevLett.129.041101",
    journal = "Phys. Rev. Lett.",
    volume = "129",
    number = "4",
    pages = "041101",
    year = "2022"
}

@article{LiteBIRD:2022cnt,
    author = "Allys, E. and others",
    collaboration = "LiteBIRD",
    title = "{Probing Cosmic Inflation with the LiteBIRD Cosmic Microwave Background Polarization Survey}",
    eprint = "2202.02773",
    archivePrefix = "arXiv",
    primaryClass = "astro-ph.IM",
    doi = "10.1093/ptep/ptac150",
    journal = "PTEP",
    volume = "2023",
    number = "4",
    pages = "042F01",
    year = "2023"
}

@article{Herman:2022fau,
    author = "Herman, Nicolas and Lehoucq, L{\'e}onard and F{\'{u}}zfa, Andr{\'e}",
    title = "{Electromagnetic antennas for the resonant detection of the stochastic gravitational wave background}",
    eprint = "2203.15668",
    archivePrefix = "arXiv",
    primaryClass = "gr-qc",
    doi = "10.1103/PhysRevD.108.124009",
    journal = "Phys. Rev. D",
    volume = "108",
    number = "12",
    pages = "124009",
    year = "2023"
}

@article{Lozanov:2022yoy,
    author = "Lozanov, Kaloian D. and Takhistov, Volodymyr",
    title = "{Enhanced Gravitational Waves from Inflaton Oscillons}",
    eprint = "2204.07152",
    archivePrefix = "arXiv",
    primaryClass = "astro-ph.CO",
    reportNumber = "IPMU22-0016, KEK-QUP-2023-0008, KEK-TH-2518, KEK-Cosmo-0311",
    doi = "10.1103/PhysRevLett.130.181002",
    journal = "Phys. Rev. Lett.",
    volume = "130",
    number = "18",
    pages = "181002",
    year = "2023"
}

@article{Mahbub:2023faw,
    author = "Mahbub, Rafid and Mishra, Swagat S.",
    title = "{Oscillon formation from preheating in asymmetric inflationary potentials}",
    eprint = "2303.07503",
    archivePrefix = "arXiv",
    primaryClass = "astro-ph.CO",
    doi = "10.1103/PhysRevD.108.063524",
    journal = "Phys. Rev. D",
    volume = "108",
    number = "6",
    pages = "063524",
    year = "2023"
}

@article{Lozanov:2023aez,
    author = "Lozanov, Kaloian D. and Sasaki, Misao and Takhistov, Volodymyr",
    title = "{Universal gravitational wave signatures of cosmological solitons}",
    eprint = "2304.06709",
    archivePrefix = "arXiv",
    primaryClass = "astro-ph.CO",
    reportNumber = "IPMU23-0008, KEK-QUP-2023-0006, KEK-TH-2514, KEK-Cosmo-0308,
  YITP-23-44",
    doi = "10.1088/1475-7516/2025/01/094",
    journal = "JCAP",
    volume = "01",
    pages = "094",
    year = "2025"
}

@article{Garcia:2023dyf,
    author = "Garcia, Marcos A. G. and Gross, Mathieu and Mambrini, Yann and Olive, Keith A. and Pierre, Mathias and Yoon, Jong-Hyun",
    title = "{Effects of fragmentation on post-inflationary reheating}",
    eprint = "2308.16231",
    archivePrefix = "arXiv",
    primaryClass = "hep-ph",
    reportNumber = "UMN--TH--4223/23, FTPI--MINN--23/15, DESY-23-122",
    doi = "10.1088/1475-7516/2023/12/028",
    journal = "JCAP",
    volume = "12",
    pages = "028",
    year = "2023"
}

@article{Kawasaki:2023rfx,
    author = "Kawasaki, Masahiro and Murai, Kai",
    title = "{Enhancement of gravitational waves at Q-ball decay including non-linear density perturbations}",
    eprint = "2308.13134",
    archivePrefix = "arXiv",
    primaryClass = "astro-ph.CO",
    reportNumber = "TU-1208",
    doi = "10.1088/1475-7516/2024/01/050",
    journal = "JCAP",
    volume = "01",
    pages = "050",
    year = "2024"
}

@article{Lozanov:2023knf,
    author = "Lozanov, Kaloian D. and Sasaki, Misao and Takhistov, Volodymyr",
    title = "{Universal gravitational waves from interacting and clustered solitons}",
    eprint = "2309.14193",
    archivePrefix = "arXiv",
    primaryClass = "astro-ph.CO",
    reportNumber = "IPMU23-0033, YITP-23-115, KEK-QUP-2023-0022, KEK-TH-2555,
  KEK-Cosmo-0325",
    doi = "10.1016/j.physletb.2023.138392",
    journal = "Phys. Lett. B",
    volume = "848",
    pages = "138392",
    year = "2024"
}

@article{Pearce:2023kxp,
    author = "Pearce, Matthew and Pearce, Lauren and White, Graham and Balazs, Csaba",
    title = "{Gravitational wave signals from early matter domination: interpolating between fast and slow transitions}",
    eprint = "2311.12340",
    archivePrefix = "arXiv",
    primaryClass = "astro-ph.CO",
    doi = "10.1088/1475-7516/2024/06/021",
    journal = "JCAP",
    volume = "06",
    pages = "021",
    year = "2024"
}

@article{Fernandez:2023ddy,
    author = "Fernandez, Nicolas and Foster, Joshua W. and Lillard, Benjamin and Shelton, Jessie",
    title = "{Stochastic Gravitational Waves from Early Structure Formation}",
    eprint = "2312.12499",
    archivePrefix = "arXiv",
    primaryClass = "astro-ph.CO",
    reportNumber = "MIT-CTP/5660",
    doi = "10.1103/PhysRevLett.133.111002",
    journal = "Phys. Rev. Lett.",
    volume = "133",
    number = "11",
    pages = "111002",
    year = "2024"
}

@article{He:2023xoh,
    author = "He, Yutong and Giri, Sambit K. and Sharma, Ramkishor and Mtchedlidze, Salome and Georgiev, Ivelin",
    title = "{Inverse Gertsenshtein effect as a probe of high-frequency gravitational waves}",
    eprint = "2312.17636",
    archivePrefix = "arXiv",
    primaryClass = "astro-ph.CO",
    reportNumber = "NORDITA-2023-066",
    doi = "10.1088/1475-7516/2024/05/051",
    journal = "JCAP",
    volume = "05",
    pages = "051",
    year = "2024"
}

@article{Shafi:2024jig,
    author = "Shafi, Mohammed and Copeland, Edmund J. and Mahbub, Rafid and Mishra, Swagat S. and Basak, Soumen",
    title = "{Formation and decay of oscillons after inflation in the presence of an external coupling. Part I. Lattice simulations}",
    eprint = "2406.00108",
    archivePrefix = "arXiv",
    primaryClass = "hep-ph",
    doi = "10.1088/1475-7516/2024/10/082",
    journal = "JCAP",
    volume = "10",
    pages = "082",
    year = "2024"
}

@article{Domenech:2024wao,
    author = {Dom{\`e}nech, Guillem and Tr{\"a}nkle, Jan},
    title = "{From formation to evaporation: Induced gravitational wave probes of the primordial black hole reheating scenario}",
    eprint = "2409.12125",
    archivePrefix = "arXiv",
    primaryClass = "gr-qc",
    doi = "10.1103/PhysRevD.111.063528",
    journal = "Phys. Rev. D",
    volume = "111",
    number = "6",
    pages = "063528",
    year = "2025"
}

@article{Jia:2024fmo,
    author = "Jia, Tianyu and Sang, Yu and Zhang, Xue",
    title = "{Nonlinear dynamics of oscillons and transients during preheating after single field inflation}",
    eprint = "2409.04046",
    archivePrefix = "arXiv",
    primaryClass = "astro-ph.CO",
    doi = "10.1103/PhysRevD.111.083531",
    journal = "Phys. Rev. D",
    volume = "111",
    number = "8",
    pages = "083531",
    year = "2025"
}

@article{Sui:2024grm,
    author = "Sui, Xiao-Bin and Liu, Jing and Cai, Rong-Gen",
    title = "{Enhancement of small-scale induced gravitational waves from solitons and oscillons}",
    eprint = "2412.08057",
    archivePrefix = "arXiv",
    primaryClass = "astro-ph.CO",
    doi = "10.1103/d2rp-xyzt",
    journal = "Phys. Rev. D",
    volume = "111",
    number = "12",
    pages = "123503",
    year = "2025"
}

@article{Aggarwal:2025noe,
    author = "Aggarwal, Nancy and others",
    title = "{Challenges and Opportunities of Gravitational Wave Searches above 10 kHz}",
    eprint = "2501.11723",
    archivePrefix = "arXiv",
    primaryClass = "gr-qc",
    reportNumber = "CERN-TH-2025-014, DESY-25-007",
    month = "1",
    year = "2025",
    journal=""
}

@article{ACT:2025fju,
    author = "Louis, Thibaut and others",
    collaboration = "Atacama Cosmology Telescope",
    title = "{The Atacama Cosmology Telescope: DR6 power spectra, likelihoods and {\ensuremath{\Lambda}}CDM parameters}",
    eprint = "2503.14452",
    archivePrefix = "arXiv",
    primaryClass = "astro-ph.CO",
    reportNumber = "FERMILAB-PUB-25-0071-PPD",
    doi = "10.1088/1475-7516/2025/11/062",
    journal = "JCAP",
    volume = "11",
    pages = "062",
    year = "2025"
}

@article{ACT:2025tim,
    author = "Calabrese, Erminia and others",
    collaboration = "Atacama Cosmology Telescope",
    title = "{The Atacama Cosmology Telescope: DR6 constraints on extended cosmological models}",
    eprint = "2503.14454",
    archivePrefix = "arXiv",
    primaryClass = "astro-ph.CO",
    reportNumber = "FERMILAB-PUB-25-0157-PPD",
    doi = "10.1088/1475-7516/2025/11/063",
    journal = "JCAP",
    volume = "11",
    pages = "063",
    year = "2025"
}

@article{Pearce:2025ywc,
    author = "Pearce, Matthew and Pearce, Lauren and White, Graham and Bal{\'a}zs, Csaba",
    title = "{Using gravitational wave signals to disentangle early matter dominated epochs}",
    eprint = "2503.03101",
    archivePrefix = "arXiv",
    primaryClass = "astro-ph.CO",
    doi = "10.1088/1475-7516/2025/11/004",
    journal = "JCAP",
    volume = "11",
    pages = "004",
    year = "2025"
}

@article{Choudhury:2025vso,
    author = "Choudhury, Sayantan and Bauyrzhan, Gulnur and Singh, Swapnil Kumar and Yerzhanov, Koblandy",
    title = "{What new physics can we extract from inflation using the ACT DR6 and DESI DR2 Observations?}",
    eprint = "2506.15407",
    archivePrefix = "arXiv",
    primaryClass = "astro-ph.CO",
    month = "6",
    year = "2025",
    journal=""
}

@article{Lynker:2025wyc,
    author = "Lynker, Monika and Schimmrigk, Rolf",
    title = "{ACT Implications for Hilltop Inflation}",
    eprint = "2507.15076",
    archivePrefix = "arXiv",
    primaryClass = "astro-ph.CO",
    month = "7",
    year = "2025",
    journal=""
}

@article{Li:2025ioq,
    author = "Li, Siyao and Yamaguchi, Masahide and Zhang, Ying-li",
    title = "{Decay and lifetime of oscillons coupled to an external scalar field: insights from instability band analysis}",
    eprint = "2507.13276",
    archivePrefix = "arXiv",
    primaryClass = "hep-ph",
    doi = "10.1088/1475-7516/2025/12/028",
    journal = "JCAP",
    volume = "12",
    pages = "028",
    year = "2025"
}

@article{Inomata:2025wiv,
    author = "Inomata, Keisuke and Kohri, Kazunori and Terada, Takahiro",
    title = "{The poltergeist mechanism -- Enhancement of scalar-induced gravitational waves with early matter-dominated era}",
    eprint = "2511.07266",
    archivePrefix = "arXiv",
    primaryClass = "astro-ph.CO",
    month = "11",
    year = "2025",
    journal = ""
}

@article{Li:2025xtf,
    author = "Li, Siyao",
    title = "{Oscillon decay via parametric resonance: the case of three-point scalar interactions}",
    eprint = "2511.03501",
    archivePrefix = "arXiv",
    primaryClass = "hep-ph",
    month = "11",
    year = "2025",
    journal=""
}

\end{document}